\PassOptionsToPackage{table}{xcolor}

\documentclass[sigconf]{acmart}
\usepackage{subcaption}
\usepackage{acmart-taps}
\usepackage{amsmath}
\usepackage{algorithm}
\usepackage{cleveref}

\usepackage{colortbl} 

\newcounter{guideline}
\renewcommand{\theguideline}{\arabic{guideline}}

\newenvironment{designbox}{%
  \refstepcounter{guideline}%
  \aptLtoX{%
    \par\medskip\noindent\textbf{Design Guideline~\theguideline}\par\smallskip
    \begin{quote}
  }{%
    \par\vspace{1em}%
    \noindent
    \begin{tabular}{@{}p{\dimexpr\linewidth-2\fboxsep-2\fboxrule\relax}@{}}
      \cellcolor{gray!99}\textcolor{white}{\textbf{Design Guideline~\theguideline}}\\
      \noalign{\vskip-\arrayrulewidth}
      \cellcolor{gray!15}%
  }%
}{%
  \aptLtoX{%
    \end{quote}\par\medskip
  }{%
      \\
    \end{tabular}%
    \par\vspace{1em}%
  }%
}

\usepackage{tabularx} 
\usepackage{multirow} 
\usepackage{enumitem} 
\usepackage{array}
\usepackage{tabularray}

\usepackage[normalem]{ulem}

\newcommand{\new}[1]{#1}
\newcommand{\remove}[1]{}
\newcommand{\note}[1]{}

\def\instructions{\instructions}
\def\instructions{\textcolor{red}}

\def\revisit{\revisit}
\def\revisit{\textcolor{blue}}

\newcommand{\diymod}{\texttt{DIY-MOD}}

\AtBeginDocument{%
  \providecommand\BibTeX{{%
    \normalfont B\kern-0.5em{\scshape i\kern-0.25em b}\kern-0.8em\TeX}}}

\copyrightyear{2026}
\acmYear{2026}
\setcopyright{cc}
\setcctype{by}
\acmConference[CHI '26]{Proceedings of the 2026 CHI Conference on Human Factors in Computing Systems}{April 13--17, 2026}{Barcelona, Spain}
\acmBooktitle{Proceedings of the 2026 CHI Conference on Human Factors in Computing Systems (CHI '26), April 13--17, 2026, Barcelona, Spain}
\acmDOI{10.1145/3772318.3790495}
\acmISBN{979-8-4007-2278-3/2026/04}


\begin{document}
\setlength{\intextsep}{5pt plus 4pt}   



\title{What If Moderation Didn’t Mean Suppression? A Case for Personalized Content Transformation}

\author{Rayhan Rashed}
\email{rayrash@umich.edu}
\affiliation{%
  \institution{University of Michigan}
  \city{Ann Arbor}
  \state{Michigan}
  \country{USA}
}

\author{Farnaz Jahanbakhsh}
\email{farnaz@umich.edu}
\affiliation{%
  \institution{University of Michigan}
  \city{Ann Arbor}
  \state{Michigan}
  \country{USA}
}

\definecolor{grey}{HTML}{969696}
\definecolor{violet}{HTML}{756bb1}
\definecolor{dgrey}{HTML}{01665e}
\definecolor{lgrey}{HTML}{5ab4ac}
\definecolor{dgreen}{HTML}{005a32}
\definecolor{purple}{HTML}{ae017e}
\def\greybar#1{{\color{grey}\rule{#1pc}{6pt}}}
\definecolor{editCol}{rgb}{0, 0, 0}
\definecolor{editColCR}{rgb}{0, 0, 0}
\definecolor{brickred}{HTML}{f03b20}
\definecolor{improveCol}{HTML}{253494}
\definecolor{worsenCol}{HTML}{d7191c}
\definecolor{violet}{HTML}{8A2BE2}
\definecolor{dodgerblue}{HTML}{1E90FF}
\definecolor{deeporange}{HTML}{EE9A00}
\definecolor{deepgreen}{HTML}{008B00}
\definecolor{mediumyellow}{HTML}{FFD900}
\definecolor{trColor}{HTML}{d01c8b}
\definecolor{ctColor}{HTML}{4dac26}

\newcommand{\checksquare}{\ding{113}}

\newcommand*{\teanna}[1]{\textbf{\sffamily{\textcolor{dodgerblue}{[#1 -- Anna Sims]}}}}
\newcommand{\farnaz}[1]{\textbf{\sffamily{\textcolor{deepgreen}{[#1 -- Farnaz J.]}}}}
\newcommand{\rayhan}[1]{\textbf{\sffamily{\textcolor{violet}{[#1 -- Rayhan]}}}}

\newcommand{\rebut}[1]{{#1}}

\newcommand{\revisioncolor}{editCol}

\newenvironment{revise}{\color{\revisioncolor}}{}

\definecolor{deepgreen}{HTML}{008B00}
\newcommand{\revisioncolorfinal}{editCol}
\newenvironment{finalrev}{\color{\revisioncolorfinal}}{}

\renewcommand{\shortauthors}{Rayhan Rashed and Farnaz Jahanbakhsh}

\begin{abstract}

Centralized content moderation paradigm both falls short and overreaches: 1) it fails to account for the subjective nature of harm, and 2) it acts with blunt suppression in response to content deemed harmful, even when such content can be salvaged. We first investigate this through formative interviews, documenting how seemingly benign content becomes harmful due to individual life experiences. Based on these insights, we developed \diymod{}, a browser extension that operationalizes a new paradigm: \emph{personalized content transformation}. Operating on a user's own definition of harm, \diymod{} transforms sensitive elements within content in real-time instead of suppressing the content itself. The system selects the most appropriate transformation for a piece of content from a diverse palette---from obfuscation to artistic stylizing---to match the user's specific needs while preserving the content's informational value. Our two user studies demonstrate that this approach increases users' sense of agency and safety, enabling them to engage with content and communities they previously needed to avoid.

\end{abstract}

\begin{teaserfigure}
    \centering
    \includegraphics[width=\textwidth]{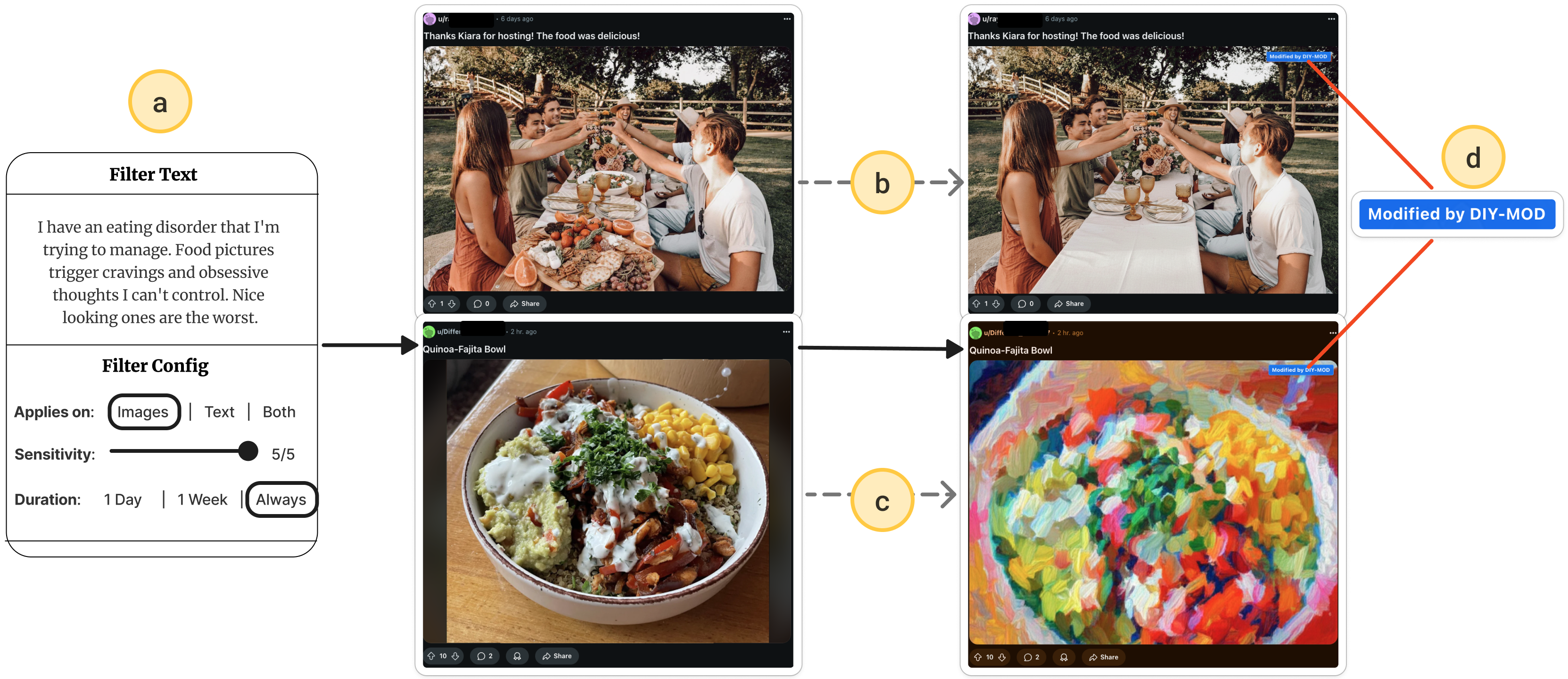}
    \caption{\diymod{} enables personalized content transformation based on individual sensitivities. A user recovering from ``eating disorder'' creates a filter (a) specifying that food imagery triggers harmful thoughts. When browsing Reddit's original feed (center), \diymod{} identifies matching posts and applies context-appropriate transformations (right): (b) semantic inpainting, which preserves the social dining context while obscuring food details; and (c) stylistic alteration, which renders the food bowl as an impressionist style painting to obscure details. A transparency indicator (d) marks all modified content. The system transforms only content matching the user's filters while leaving other posts unchanged.}
    \label{fig:teaser_user_pov}
\end{teaserfigure}


\maketitle

\section{Introduction}
\label{sec:intro}

\begin{quote}
\textbf{Aisha}\footnotemark[1]{}, a mother still reeling from a recent miscarriage, scrolls through her social media feed. It's a mindless, everyday act. Then, suddenly, it appears: a pregnancy announcement from a friend. For most, it’s a moment of shared joy. But for Aisha, the experience is devastating. An unexpected wave of grief washes over her, immediately followed by a single, painful question: \textit{Why them, and not me?} This innocent post, meant to celebrate new life, lands as an unexpected blow. Aisha knows this isn't her friend's fault and this is likely the first of many updates. Yet the platform offers no mechanism to filter such deeply personal sensitivities. Her options here are stark: unfollow her friend and sever a valued connection, or remain and endure repeated distress.
\end{quote}
\vspace{3pt}
\begin{quote}
\textbf{Rex}\footnotemark[2]{} faces a different struggle. Recovering from binge-eating disorder, he finds that food images trigger intense cravings and obsessive thoughts he cannot control. Yet his feed overflows with food content---every restaurant visit, every home-cooked meal, every dessert becomes a photo opportunity. Friends pose with loaded plates at gatherings, weaving food throughout the social moments he values: seeing who attended which event, staying connected to his community. He wants to see his friends, but not the food that dominates every frame. He has already blocked several accounts when it became overwhelming. But how many can he block when everyone shares food?
\footnotetext[1]{\textsuperscript{, 2}~Aisha and Rex are pseudonyms for two of our study participants. The vignettes are based on the experiences they shared with us during our interview, and we have received their explicit permission to use their story.}
\end{quote}

\noindent These two scenarios highlight a fundamental challenge in online content moderation, defined by two interconnected problems. First, what counts as harmful varies drastically by user, over time, and across situations. Centralized, one-size-fits-all policies, designed to filter broadly proscribed content like hate speech or graphic violence, struggle to accommodate the deeply personal and contextual nature of content sensitivity \cite{roberts2019, gillespie2018}. Second, existing moderation tools, either those exercised by the platforms or given to users, rely on blunt interventions that effectively boil down to suppression of whole posts or accounts---whether through removal, algorithmic downranking that buries them out of sight, muting all posts containing certain keywords, or unfollowing. These interventions force users into the impossible trade-off Aisha and Rex face. These blunt instruments cannot differentiate between a single triggering element and the valuable surrounding content. The result is a choice between enduring harm and complete disconnection.

This paper argues for and explores a new direction for moderation that addresses both problems: \textit{personalized content transformation}. This paradigm treats the experiential layer of content as something that can be selectively transformed in ways that support user safety while preserving informational value. Such an approach could allow users to remain engaged with their communities and access valuable information while mitigating exposure to aspects of content they find distressing.

Building on insights and design guidelines from our formative study (\Cref{sec:method-1}) with 12 participants experiencing diverse personal sensitivities, including phobias, PTSD, and beyond, we designed and developed \textbf{\diymod{}} (\texttt{\textbf{D}o-\textbf{I}t-\textbf{Y}ourself \textbf{Mod}eration}), a browser extension that operationalizes the concept of personalized content transformation (\Cref{sec:system_architecture}). \diymod{} empowers users to create their own moderation layer on top of mainstream platforms. It moves beyond binary show-or-hide filtering to offer a palette of nuanced, multi-modal interventions. Users can have a dialogue with \diymod{} in natural language to describe what they wish to avoid. In response, \diymod{} transforms elements in content that match user sensitivities while preserving surrounding context and meaning. For visual content, this includes making certain elements more abstract, reducing realism through artistic rendering, or applying visual euphemisms---modifying triggers while keeping the overall presentation and semantics faithful to the original. For textual content, this includes regional blurring to rewriting the text.

We evaluate \diymod{} through two user studies. Our first study (\Cref{sec:user-study-1}), an in-situ evaluation of naturalistic browsing on Reddit feeds, finds that personalized transformation increases users' sense of agency and enables them to safely engage with content they would otherwise avoid. \remove{Our second study (\Cref{sec:user-study-2}), } \remove{a controlled} \new{However, because real-world browsing offers sporadic and unpredictable exposure to triggers, we conduct a second, controlled study (\Cref{sec:user-study-2}) to compare transformation techniques.}  \new{This} preference elicitation identifies a set of principles for effective content transformation, such as the need to provide cognitive closure. Together, our findings demonstrate that this approach offers a promising and user-valued alternative to suppressive content moderation. 
In summary, this paper makes the following contributions:
\begin{enumerate}
\item An empirical account of subjective harm online, detailing the failure of the status-quo centralized moderation paradigm through a formative study with 12 participants (\Cref{sec:method-1}).

\item The design and implementation of \diymod{}, a working browser extension that operationalizes personalized content transformation
(\Cref{sec:system_architecture}).

\item An in-situ evaluation of \diymod{} demonstrating that personalized transformation increases users' sense of agency and safety during real-world use (\Cref{sec:user-study-1}).

\item A controlled study that contributes further principles for designing user-aligned transformations, such as the need to provide cognitive closure (\Cref{sec:user-study-2}).
\end{enumerate}


\section{Related Work}

Our work stands on \textbf{three} related pillars: centralized moderation and its structural limits; evidence that experiences of harm are subjective; and systems and architectures that shift more agency to the user. Our system design is also informed by technical advances in content modification and by therapeutic practices. 

\subsection{Centralized Moderation}

Platform governance scholarship characterizes platforms as private rule-setters that curate and enforce boundaries for what their users may see~\cite{gillespie2018}. 
In practice, we refer to centralized moderation as arrangements where end-users have little to no say in defining or enforcing the rules~\cite{sankin2017activists, trustnet_user}. Instead, decisions are made by platform policy teams, automated systems, or designated intermediaries such as community or instance administrators~\cite{lee2024people, diaz2021double}. While these intermediaries represent a form of delegated authority, their power is ultimately constrained by boundaries set by the platform. 
Platforms typically seek to avoid liability for third-party content~\cite{keller2023carriage, gillespie2010politics}, but they increasingly act as setters of norms~\cite{gillespie2018} and enforcers of the rules they establish.\footnote[3]{~We take as given that clearly illegal content (e.g., CSAM~\cite{CSAM}, direct and credible incitement to violence) must be removed under law and platform policy. Our focus is lawful content that can nevertheless be distressing or harmful depending on the recipient and context.} To operate at scale, these arrangements posit a \emph{common boundary} of ``harmful'' content that can be expressed as rules and categories and applied uniformly.

At platform scale, such a shared boundary needs to fit many norms and situations~\cite{gillespie2020content, gorwa2020algorithmic, keller2023}. Fixed categories make enforcement easier and consistent but risk under-suppressing content that some decry as harmful, and over-suppressing content that some argue does not have the potential to harm~\cite{gillespie2018, lee2024people}. The underlying assumption in this approach is that content should be treated the same for everyone. But in practice, there is frequent and contentious debate on whether the policies themselves are legitimate or appropriate~\cite{douek2021governing, caplan2018content}.

A universal boundary enables platforms to hand classification to AI systems and human moderators. Yet cases that sit at the edge, or events not covered by existing rules, produce inconsistency and ad hoc decisions~\cite{myerswest2018, trustnet_ai, jahanbakhsh2023thesis}. Once flagged, enforcement usually means visibility limits such as downranking, gating, or removal~\cite{gorwa2020algorithmic}. Softer measures like content warnings (CW) still depend on centralized labels, where either the platform or the poster assigns a predefined category. Several scholars also propose applying restorative and transformative justice frameworks to platform governance as alternatives to these punitive approaches ~\cite{hasinoff2022scalability, xiao2023addressing, schoenebeck2020reimagining, xiao2022sensemaking}

A limited form of decentralized decision making exists where subcommunities or instances set their own rules~\cite{fiesler2018reddit, kuo2023unsung}, or when experiments like digital juries invite peers to deliberate on specific moderation cases~\cite{fan2020digital}. Yet these approaches still require defining universal categories of acceptable content applied to all members, regardless of context or individual differences~\cite{trustnet_user}.



Across policies, automation, human review, delegation, and CW, centralized moderation rests on \textbf{one premise}: \textit{harm can be defined with a common boundary and applied uniformly}. Next, we show how lived experiences of harm complicate this assumption.

\subsection{Harm as Situated and Subjective}


Much debate has centered on where the boundary of harmful content lies. The pursuit of a universal boundary, however, is inherently flawed: it presumes that a single set of rules can capture the diversity of how people experience harm. We build this case by not focusing on the familiar ``gray areas’' that already generate public controversy~\cite{douek2021governing, caplan2018content,diaz2021double}, but by turning to content that is rarely debated at all---content that most people overlook but that can be profoundly harmful for individuals with specific lived experiences, such as phobias or  PTSD triggers. 

Specific phobias, some with genetic roots ~\cite{kendler2001genetic,villafuerte2003genetic}, cause intense fear responses to otherwise low-risk stimuli~\cite{apa2013dsm5, nimh2022phobias}. Even digital exposure~\cite{kff_needle_phobia_article, michalowski2017set} can provoke immediate physiological reactions, including panic attacks~\cite{ost1989phobias}. These sensitivities can have societal consequences: people with trypanophobia (needle phobia) have reported that frequent needle imagery in vaccine news contributed to their vaccine hesitancy during COVID-19~\cite{kff_needle_phobia_article}.

Trauma and life events also shape sensitivity. Each year, 3–4\% of adults meet criteria for PTSD~\cite{kessler2017ptsd}. Pregnancy loss affects 10–20\% of known pregnancies; content about pregnancy or infants--benign for many--can be acutely painful for some~\cite{bardos2015miscarriage}. Eating disorders affect roughly 9\% of the population and heighten sensitivity to food, weight, and body imagery~\cite{hudson2007eating}. In all cases, both the \emph{threshold} and the \emph{interpretation} of distress vary across people and over time.

Online environments make these differences especially visible. The same post can be celebratory for one person and painful for another, depending on needs and timing. Prior work on pregnancy loss documents this divergence in social media contexts~\cite{andalibi2018announcing, andalibi2021social}. In eating-disorder contexts, people often seek support while trying to avoid content that exacerbates symptoms~\cite{pater2016hunger, chancellor2016quantifying}. A single category boundary cannot account for the range of experiences among users.
A complementary lens from law and psychology also emphasizes  subjective harm---the internal distress tied to a person's experience, identity~\cite{SHEEQ}, or perception~\cite{calo2011boundaries}. 

Together, these show how \textbf{harm experienced} by individuals can be \textit{decoupled from a common boundary}.  Content within the boundary can still cause real harm for individuals.

\subsection{User Agency and Personalized Control}

Two lines of prior work seek to give users more control and agency while still operating within a common content boundary: i) architectures that shift the boundary-setting downstream towards users, and ii) end-user facing tools that tune exposure. We outline what these approaches afford and where they fall short.

\noindent\textit{Architectural approaches:} Middleware proposals~\cite{fukuyama2021} ask platforms to expose interfaces so third parties can curate on a user's behalf; users then can choose the provider whose values align with theirs. Federated and protocol-centric designs (e.g., Mastodon\cite{mastodon}, Bluesky\cite{bluesky}) push choice to the instance or client layer, where admins or client developers set policies and content boundaries~\cite{masnick2019}. In both cases, the gatekeeping moves closer to the end-users, but \textbf{categories} and \textbf{thresholds} are still \textit{ defined upstream}; end-users pick from presets.

\noindent\textit{User-facing tools:}
Prior HCI work shows strong demand for control over one’s feed~\cite{jhaver2023personalizing, jhaver2023users, trustnet_user}. 
Grassroots systems often layer community-generated safety signals directly onto existing platform interfaces. A prominent example is Shinigami Eyes: community labeling that marks spaces as welcoming or hostile to certain marginalized groups~\cite{shinigami-eyes-2024, devito2023shinigami, saetra2023shinigami}. Trust-based frameworks enable users to delegate judgments of content accuracy to selected trusted peers or services~\cite{trustnet_browser, trustnet_user, trustnet_ai}. At scale, these trust-based frameworks create collaborative signal pipelines that leverage the collective judgment of the community. Systems such as Crossmod~\cite{eshwar_crossmods}, help early detection of problematic comments that would be removed by moderators.  These tools increase transparency and participation, but their \textbf{mechanism} \textit{remains suppressive of the entire content}.
 
\label{sec:related-work}
\subsection{Towards Personalized Content Transformation}
Status-quo moderation approaches operate at the post level, adjusting its visibility through removal, downranking, or limited distribution. These interventions affect the entire post regardless of whether it contains both harmful and valuable elements. User-side controls similarly filter exposure to entire posts or accounts through blocking, muting, or ``show me less of this'' feature. \new{However, these controls have social costs. Blocking can appear hostile or signal `defeat', which harassers treat as a badge of honor~\cite{heung2025ignorance}. By severing ties completely~\cite{jhaver2018online}, it forces a choice between enduring harm and losing social context and threat-monitoring capabilities.}

We take a different path with \emph{content transformation}, drawing on exposure therapy’s principle of safe engagement with weakened stimuli such as photos~\cite{rothbaum2002exposure, yoshikawa2024digital, foa2007trigger}. \new{Clinical variants, such as exposure and response prevention (ERP) for obsessive–compulsive disorder~\cite{bjorgvinsson2007obsessive, abramowitz2019exposure}, use this principle in structured therapeutic programs. Recent work observes that triggers and tolerances both vary across people and change over time, and accordingly recommends that systems supporting ERP be adaptive~\cite{wang2025mentally}.} Prior HCI work has similarly modified visual content to protect privacy while preserving utility via obfuscation~\cite{li2022_obfuscation, monteiro2024_obfuscate, hasan2019_obfuscate} and cartoonification~\cite{hasan2017cartoonin_obfuscate}. We apply this principle in a non-clinical,
everyday setting: \textit{minimizing harm while maintaining the post's informational and social value.}

Recent AI advances provide the foundation for personalized transformation. Multimodal models (LLMs~\cite{devlin2019bert, kumar2023llm} and VLMs~\cite{zhu2023minigpt, liu2024improved}) distinguish central from peripheral content elements, enabling targeted interventions. Generative models then execute transformations: diffusion models remove or replace visual elements~\cite{rombach2022high-resolution-diffusion}, style-transfer changes presentation~\cite{gatys2016image-style-transfer}, and language models can rewrite text while preserving meaning. Together, these tools allow selective modification of distressing elements while retaining core information. Because harm is subjective, the same piece of content can be transformed in several plausible ways. The challenge is choosing which best serves a given user. We adapt the LLM-as-a-judge approach~\cite{zheng2023judging, liu2023gpteval} to evaluate transformations through a personalized lens. The model receives both the transformed content and detailed user context
to predict which intervention would best balance reducing harm with preserving information for that individual.



\section{Formative Study}
\label{sec:method-1}

To ground our system design in the lived experiences of users, we conducted a formative qualitative study. Our goals were to map the long tail of content that individuals experience as harmful, document how people cope with such distressing encounters online, and surface limits in current tools while eliciting requirements for personalized transformation. The complete study protocol was approved by our institution's IRB.

\subsection{Methods}
\subsubsection*{Participant Recruitment}

We recruited participants between December 2024 and January 2025 by posting our study invitation to university-affiliated mailing lists and various online communities where we expected we might find users whose needs are unmet by status-quo moderation schemes, including subreddits related to PTSD and phobias as well as mental health forums~\footnote[4]{~We obtained permission from the moderators of all these communities before posting.}. The recruitment materials invited individuals who had ``encountered anxiety-inducing, unsettling, or unwanted content while browsing online.'' Interested individuals completed a screening survey to assess initial eligibility. The survey confirmed that participants were 18 years of age or older and, as stipulated by our IRB due to international data privacy regulations, were not located in the UK or EU.

Our screening criteria focused on participants who could recall recent, concrete instances where otherwise ordinary content felt distressing because of personal context (e.g., phobias, trauma, troubling life experiences, or abuse), regardless of how platforms currently categorize or moderate such content. We recruited twelve participants whose experiences challenge conventional assumptions about harmful content. Their sensitivities spanned specific phobias (e.g., Kosmemophobia, Arachnophobia), trauma-related triggers (PTSD), and life events (e.g., pregnancy loss)---content ranging from what platforms ignore entirely to what they actively moderate. Following established practice of designing for `extra-ordinary' users to improve systems for all~\cite{extra_ordinary_user}, we began with these cases. But we note that these experiences are not fundamentally different from what others may encounter. Content sensitivities can arise for anyone at different points in life. Upon completion of the full interview, each participant received a \$25 USD gift card.

\subsubsection*{Interview Protocol}

After giving informed consent, participants joined a 60-minute semi-structured interview on Zoom in English. We asked them to describe personal encounters with distressing or harmful online content, reflect on the effectiveness of existing moderation tools, and talk about how the form of content--whether text, images, or video--shaped its impact. We also asked their perceptions about a hypothetical moderation tool that allowed them to personalize their content boundaries and incorporated our envisioned mechanisms like content transformation. 

Recognizing the sensitive nature of the topics, we explicitly informed participants of their right to pause, skip any question, or end the interview at any time. The consent form also provided contact information for mental health support resources. Following the interview, participants completed a brief post-interview questionnaire to provide demographic information.

\label{sec:findings-interview}
\subsection{Findings Overview}\label{findings:severity}\label{findings:begin}

Our interviews revealed that individuals encounter content online that causes psychological harm, requiring constant vigilance and affecting their well-being. Understanding these lived experiences is essential for designing effective moderation systems. Throughout the findings for our formative study, where we present participants' free-text responses, we identify them with a string of the form ``P\_0\_'' + an identifier to preserve their anonymity.

Reactions to sensitive content can be intensely physical. P\_0\_04, who has a severe phobia of centipedes, described breaking their phone after involuntarily throwing it upon seeing an image: \textit{``I react before I even realize I've really seen it. It's just so fast.''} This immediate response illustrates the visceral nature of sensitivities to digital content. For others, exposure causes prolonged psychological disruption,
for instance needing to lie down and practice breathing exercises to regain stability.

Triggers often extend beyond simple, concrete objects to entire discourse categories. Avoiding them requires filtering broad swaths of content. P\_0\_08, a military veteran with PTSD, finds any political content acting as a powerful trigger due to its connection to his past service experience where he was \textit{``ordered to stand down and watch as they [civilians] were attacked.''}

\subsection{The Nature of Personal Content Sensitivities}

\subsubsection{A Spectrum of Triggers: From Phobias to Values}
\label{subsec:findings_granularity}

Beyond the relatively ``easy'' cases where restrictions are broadly accepted~\cite{schoenebeck2023online,klonick2017new}, much of the tension over what should or should not be allowed in online spaces stems from the legitimate subjectivity of harm.

For example, personal sensitivity can sometimes be intertwined with a significant life event. For P\_0\_12, the period following a miscarriage transformed otherwise joyous content like pregnancy announcements from friends into sources of profound pain.
This challenge intensifies when the distressing content is socially ubiquitous and celebrated. P\_0\_03, who has Kosmemophobia (phobia of jewelry), described the near impossibility of avoidance:
\textit{``because all kinds of photos all the time show up everywhere--TikTok, Reddit, anywhere.''}
Beyond phobias, content sensitivities can stem from lived trauma, such as P\_0\_08's reaction to political content, or from conflicts with ethical principles. P\_0\_09's veganism transforms everyday food advertisements into sources of genuine distress. She explained that because consumption of animal products is ``so normalized'', she perceives cruelty in images that appear innocuous to the general public.
These findings demonstrate that content harmless to many can be harmful to some, calling into question the pursuit of a platform-wide consensus boundary for moderation.

\subsubsection{The Nuance of a Trigger: Granularity and Context-Dependence}
\label{subsubsec:nuance-triggers}

Participants described sensitivities that hinged on fine-grained details. For P\_0\_03, her Kosmemophobia was not uniform, but varied by material and presentation:

\begin{quote}
``The piercings are definitely the worst.[...] Touching the skin in general is uncomfortable. It just sort of feels gross and dirty in some ways. And metal. The material also tends to be a problem.''
\end{quote}

\noindent This sensitivity sometimes extends to unrelated objects sharing similar visual patterns which then activate the same response. P\_0\_04 described experiencing her centipede phobia when seeing brain coral in her son's ocean book because the coral's pattern resembled centipede shapes.

\begin{designbox}
Enable users to define content sensitivities at their chosen level of specificity rather than constraining them to predetermined classification schemes.
\end{designbox}
\subsection{The Failure of Platform-Level Moderation}
\label{subsec:findings:failure-of-platforms}
\subsubsection{When Platform Tools Fail and Lose Trust}

Participants described platform-level tools as ineffective and misaligned with their nuanced needs.
Generic warnings were too broad to be actionable. P\_0\_04 described Reddit’s ``Not Safe For Work'' (NSFW) tag as unhelpfully vague: \textit{``It could be anything ... you just don't know.''}
\noindent
Algorithmic knobs also underdelivered. P\_0\_03 found that such knobs fall short of their promise:
\begin{quote}
``I really don't use the don't show me this, or show me less like this on TikTok, because it has never been effective for me, and in my experience tends to only just make it show me more of the same content.''
\end{quote}
\noindent
Participants emphasized that these controls amount to \emph{teaching the algorithm}, not protection. The signals are unclear and the results are not immediate: users must guess which actions matter, provide many examples, and wait for the model to adjust. As P\_0\_02 put it, \textit{``it won’t actually stop immediately--you have to give them a lot of time.''} Another participant noted that such knobs gives little feedback about what changed or why.

These experiences eroded confidence in platform-led moderation tools. P\_0\_04 stated simply: \textit{``I don't trust in the platform at this point to do it accurately.''}
This distrust is fueled by skepticism about platform motivations. P\_0\_11 questioned platform incentives:
\begin{quote}
``It thrives off of people being upset [...] I don't think most social media [...] cares about a proper system.''
\end{quote}
\noindent
We found that for some participants, this distrust has led to complete disengagement from platforms and seeking alternatives to platform-controlled moderation.

\begin{designbox}
    Build transparent systems where users understand when content 
is filtered and can easily override decisions, building trust through control.
\end{designbox}

\subsubsection{The Demand for Personal Agency}

The universal distrust of platforms, combined with the highly personal nature of triggers, led all participants to demand direct control over their filtering decisions. P\_0\_05 articulated this as a balance between personal protection and others' freedom:
\begin{quote}
``I don't want to take away somebody else's ability to talk about the things they want to talk about. But at the same time I also don't want to have to see what they have to say if it's gonna be a problem.''
\end{quote}

\noindent Participants universally wanted to define their own content boundaries. As P\_0\_06 stated, \textit{``I'll prioritize my own need. I'm the affected person here.''} Similarly, P\_0\_08 emphasized wanting \textit{``the option to tailor it for myself.''} 

\begin{designbox}
    Implement intervention at the recipient level, rather than constraining the poster, ensuring one user's safety needs do not restrict others' expression.
\end{designbox}

\subsection{The Impossible Trade-off: Connection vs. Safety}

\subsubsection{The Blunt Instrument Problem}

A consistent theme was the trade-off users face between maintaining social connections and protecting their emotional well-being. Current tools---primarily blocking or unfollowing sources---force users to often bluntly sever valuable relationships to avoid occasional distressing content.

P\_0\_08 voiced this difficulty when discussing how he handles triggering content from otherwise valued sources: \textit{``Well, I guess I'm just gonna miss out on everything else they have to say.''}
This bluntness becomes particularly problematic when content contains both harmful and valuable elements. P\_0\_03 explains a ``filter paradox'' she faces where blocking a tag like ``epilepsy'' to avoid seizure-inducing Photos/GIFs would also inadvertently block vital support conversations for people with epilepsy. Similarly:

\begin{quote}
``If I have people tagging a phobia.[...] I might want to see people discussing the phobia and be part of that conversation as a person who also has it. But I also want to be able to block posts that are triggering to the phobia. And that is a lot harder to navigate.''
\end{quote}

Similarly, P\_0\_11, who has an eating disorder, wants to see friends gathering at restaurants---the people, the ambience, the social moments---but cannot tolerate the food that inevitably appears in these photos. He cannot filter restaurant content without losing track of his social circle's activities.
\noindent
These experiences illustrate a crucial insight: content that contains distressing element often simultaneously carries social or informational value. The binary choice of block-or-endure fails to recognize that harmful content elements may coexist with valuable ones.
\begin{designbox}
Recognize that harmful and valuable content often coexist. Design systems that 
modify the experiential layer rather than removing content entirely.
\end{designbox}

\subsection{Coping Strategies and Their Limitations}
\label{subsec:findings:coping-with-platforms}
\subsubsection{Manual Labor of Safety}
\label{subsec:findings_temporal}
In the absence of effective platform tools, participants described elaborate manual strategies for managing their online experience. P\_0\_04 described her social media use as a constant assessment of her own resilience, calling it a form of \textit{``Russian roulette''} that depends on her \textit{``risk tolerance level for the day.''}
This requires substantial ongoing effort:
\begin{quote}
``I appreciate that the whole thing is dynamic like I can re-follow people, I can re-add, I can add subreddits back in, and I do, depending on my mood.[...] although it is annoying to like manually add and remove.''
\end{quote}

For other participants, when filtering is not possible, the only remaining strategy is complete disengagement. P\_0\_12 described how, in the months following her pregnancy loss, she \textit{``didn't even go online because I was so afraid of seeing something that would be upsetting''}. 

\begin{designbox}
    Minimize the labor of safety by supporting adjustments that adapt to a user's fluctuating sensitivities.
\end{designbox}

\subsubsection{Community Care and Social Filtering}
\label{subsubsec:findings:community-care}

While some users exhaust themselves with individual coping strategies, others turn to their communities for protection. This ``social safety net'' relies on collective care, where users voluntarily take on the labor of protecting one another. P\_0\_03 described this culture on Tumblr: \textit{``My friends on Tumblr are very generous, and will tag the things that upset me. Specifically, we do this for each other.[...] And then we all help each other collectively.''}

However, this human-powered system, while effective in small, close-knit communities, is fragile and does not scale to larger platforms or casual social connections. Such systems depends entirely on others' goodwill and awareness---resources that cannot be guaranteed outside of carefully cultivated spaces.

\subsection{Context Determines Content Impact}\label{findings:end}

Participants revealed that the same content can have dramatically different impact based on presentation and context. P\_0\_09 articulated how visual presentation affect her response to animal product imagery:
\begin{quote}
``If it's an advertisement for like a burger that's like covered in sauce, you barely see it, and it's like cooked well done, then I think my brain is more able to disassociate from that, even though I know it's an animal.''
\end{quote}

P\_0\_08 found text more distressing because \textit{``the imagination is always worse than reality. Text kind of leaves it up to your imagination
to fill in the the blanks.''} Conversely, P\_0\_04 viewed text as less threatening and even helpful as an early warning system: \textit{``If I see it written, that's like tells me a message like I'm gonna get out of this subreddit.[...] there might be a picture coming.''}

When asked about hypothetical interventions to modify content, participants expressed varied preferences.
Some favored simple blurring, while others worried that visible modifications could draw more attention---what P\_0\_03 called ``knowing it's there'' problem. These varying responses to content presentation and intervention methods underscore that no single approach would work for all users. 

\begin{designbox}
Design interventions that adapt to individual differences in how 
content modality and presentation affect distress.
\end{designbox}

\subsection{Lessons Learned}
Our findings reveal \textbf{three} fundamental misalignments between how platforms approach content moderation and how users experience harm online. 
\vspace{-2pt}
\begin{enumerate}
    \item  Personal sensitivities span ethical values, trauma responses, and specific phobias, challenging any universal boundary for ``harmful'' content.

    \item These sensitivities fluctuate with state and context. Static, platform-level controls miss temporal variation.

    \item Harmful and valuable elements can coexist within the same content. When moderation suppresses entire posts, it forces users to sacrifice either safety or connection.
\end{enumerate}
\vspace{-2pt}

Participants' experiences point toward a different approach. They need to specify sensitivities in their own terms, not platform categories. They need interventions that adapt to both content characteristics and personal context. They need control and feedback without being forced into tedious workarounds. The next section describes \diymod{}, our system designed to address these needs through \textit{personalized content transformation.}



\section{The System: \diymod{}}
\label{sec:system_architecture}

We designed \diymod{} as an open source browser extension
\footnote[5]{~The source code can be found at \href{https://github.com/UMichHCI/diymod}{https://github.com/UMichHCI/diymod}}
that creates a personalized content moderation layer between users and platform content. 
Consider Rex from our opening vignette. After installing \diymod{}, he opens the extension popup. Through the conversational interface, he types: ``I have an eating disorder that I'm trying to manage. Food pictures trigger cravings and obsessive thoughts I can't control.'' 
He then configures the filter with a high sensitivity level to apply only to images, setting the duration to `Always'.

Minutes later, Rex scrolls his Reddit's popular feed: \texttt{r/popular} and encounters a post celebrating a community event, with an image showing people gathered around a table filled with food. \diymod{} intercepts the posts (text and image) before it reaches the browser. The system's backend analyzes the text and image against Rex's filter and applies an appropriate transformation, such as using semantic inpainting to remove the food while preserving the social context of the gathering (\Cref{fig:teaser_user_pov}b). A small \textsc{``Modified by DIY-MOD''} indicator appears (\Cref{fig:teaser_user_pov}d). Rex can now safely engage with the post without being exposed to his triggers. The original post remains unchanged for everyone else; the transformation happens only in Rex's browser.

This scenario illustrates \diymod{}'s \textbf{three} core components.
First, users create filters through natural language conversation or image upload using the extension popup. Second, platform adapter intercepts incoming content during natural browsing before it is handed off for rendering~\cite{piccardi2024reranking, kolluri2025alexandria}. Third, our backend analyzes content using large vision-language models (LVLMs) and applies appropriate transformations and returns instructions and data to the adapter to render locally. 
We describe them in the following sections.

\subsection{Filter Creation and Configuration}
\label{subsec:filter_create_and_config}

Users describe their sensitivities in natural language to \diymod{} using their own words or by uploading example images. To capture the user's specific needs, the system engages in conversational grounding~\cite{cho2020role, brennan2014grounding}. The underlying language model is prompted to seek clarification when a description is ambiguous or underspecified, helping the user iteratively refine the filter. For instance, after a user uploads an image of jewelry, the system might ask: \textit{``Does this sensitivity apply to all jewelry or specific types?''}  Similarly, an initial filter for political content might prompt the system to ask: \textit{``Does this apply to all political topics, or specific ones like discussions of civilian casualties?''}

This conversational grounding allows users to define sensitivities at their own level of specificity, rather than being constrained to predetermined categories \textbf{(DG1)}. It can result in nuanced filters, for example \textit{``Images of spiders, mostly close-ups; tiny or distant images are okay.''} or \textit{``Political discussions, but only when they mention civilian casualties.''} The system can then distinguish between spider proximity that matters to someone with arachnophobia, or between general political discourse and the specific contexts that trigger a veteran's PTSD.

\noindent After establishing the filter description, users access \textbf{three} configuration controls (see \Cref{appendix:diymod}):

{\textbf{(a) Sensitivity Level}}: This indicates the user's level of distress when encountering content related to the filter they are configuring. The system uses this information when deciding on what transformation is appropriate for a user and post. Less intense aversions might result in blurring words or image regions, while for more intense aversions, rewriting entire passages or overlay warnings may be more appropriate. 

{\textbf{(b) Content Modality:}} Specifies whether transformations should apply to text, images, or both. This control respects that sensitivities manifest differently across modalities. Some users find textual mentions distressing while others only react to visual depictions.

{\textbf{(c) Duration:}} Filters expire after 24 hours, one week, or never. Temporal controls let users adapt without constant manual adjustment, reducing the ``manual labor of safety''{} \textbf{(DG5)}.

The extension popup interface provides essential controls within limited screen space. Users manage their filters (e.g., custom time limits, additional metadata, or editing filter descriptions) by accessing the \emph{options page}~\cite{options_page} of the extension. When creating filters, \diymod{} can also detect potential overlaps and offer to modify existing filters.

\subsubsection*{Filter Storage and Sharing}
\diymod{} operates in two modes to accommodate different privacy needs. By default, the system runs anonymously and filters are stored locally in the browser. Users can optionally authenticate through Google to sync their settings across devices.

Both modes support filter export and import as JSON files. This reduces setup burden in two ways. First, it provides data portability for anonymous users. Second, it enables grassroots filter sharing, where users can curate sophisticated filter sets for specific sensitivities and share them within their communities. A support group member who has refined filters for eating disorder content can export their configuration for others facing similar challenges. This reduces burden for new users.

\subsection{Content Processing Client} 


\diymod{} uses a client-server architecture. The browser extension client handles interception of incoming content to the browser and rendering of modified content.

\subsubsection*{Platform Adapters and Content Interception}

The client employs platform-specific adapters that intercept network responses containing new content elements~\cite{piccardi2024reranking}. This design keeps the filtering logic platform-independent, which we validated by implementing an adapter for Reddit.
When adapters detect new posts, they extract text and image URLs before content is handed off for rendering. Only public content leaves the client and no user identifiers or profile information is transmitted to the server. The server analyzes this content against the user's filters and returns personalized transformation instructions. The adapter then applies transformations to content within the post's original layout before content becomes visible to the user. This client-side application of targeted modifications ensures transformations are applied at the recipient level \textbf{(DG3)} and preserves valuable surrounding elements within the content \textbf{(DG4)}.

\begin{figure*}[!htbp]
    \centering
    \begin{subfigure}[b]{0.32\textwidth}
        \includegraphics[width=\textwidth]{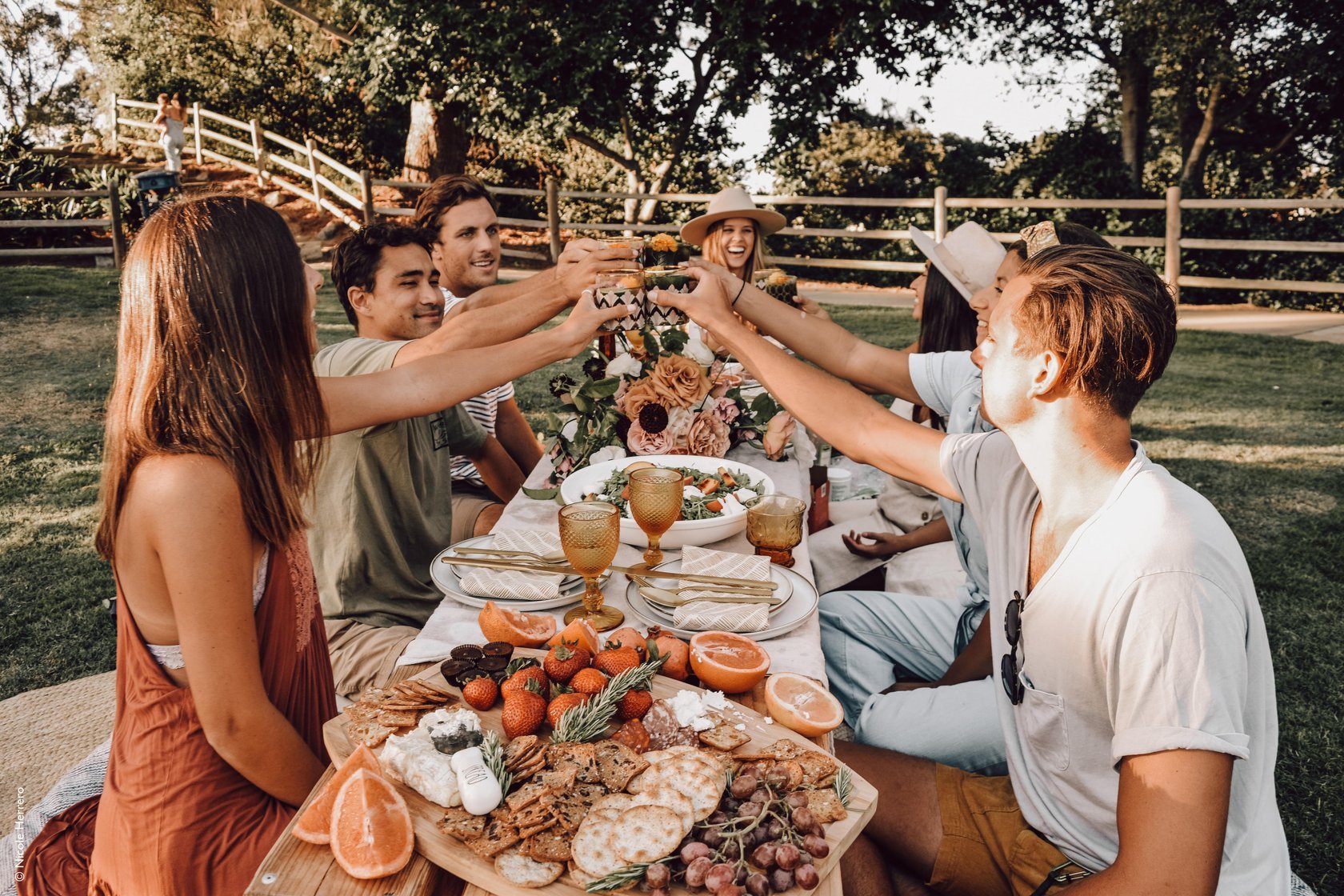}
        \caption{Original image}
        \label{fig:original}
    \end{subfigure}
    \hfill
    \begin{subfigure}[b]{0.32\textwidth}
        \includegraphics[width=\textwidth]{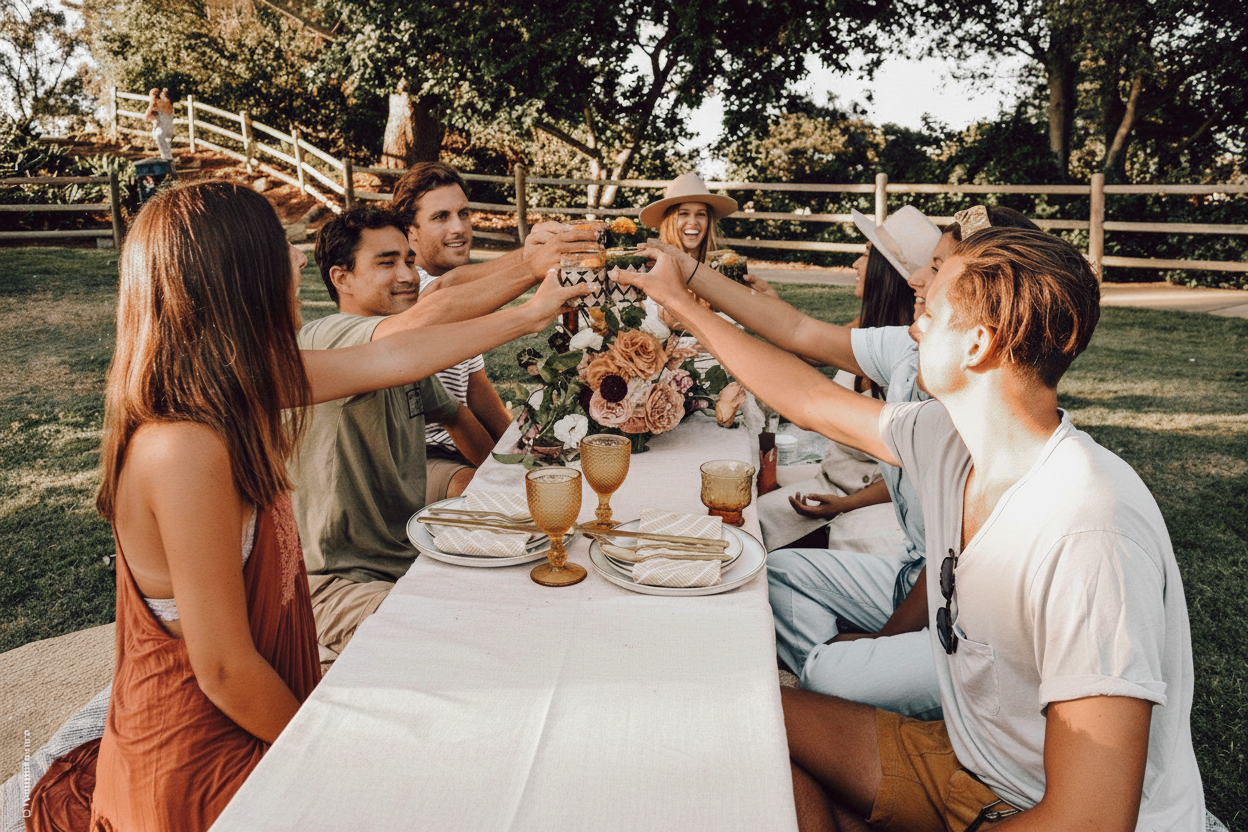}
        \caption{Inpainting}
        \label{fig:inpainting}
    \end{subfigure}
    \hfill
    \begin{subfigure}[b]{0.32\textwidth}
        \includegraphics[width=\textwidth]{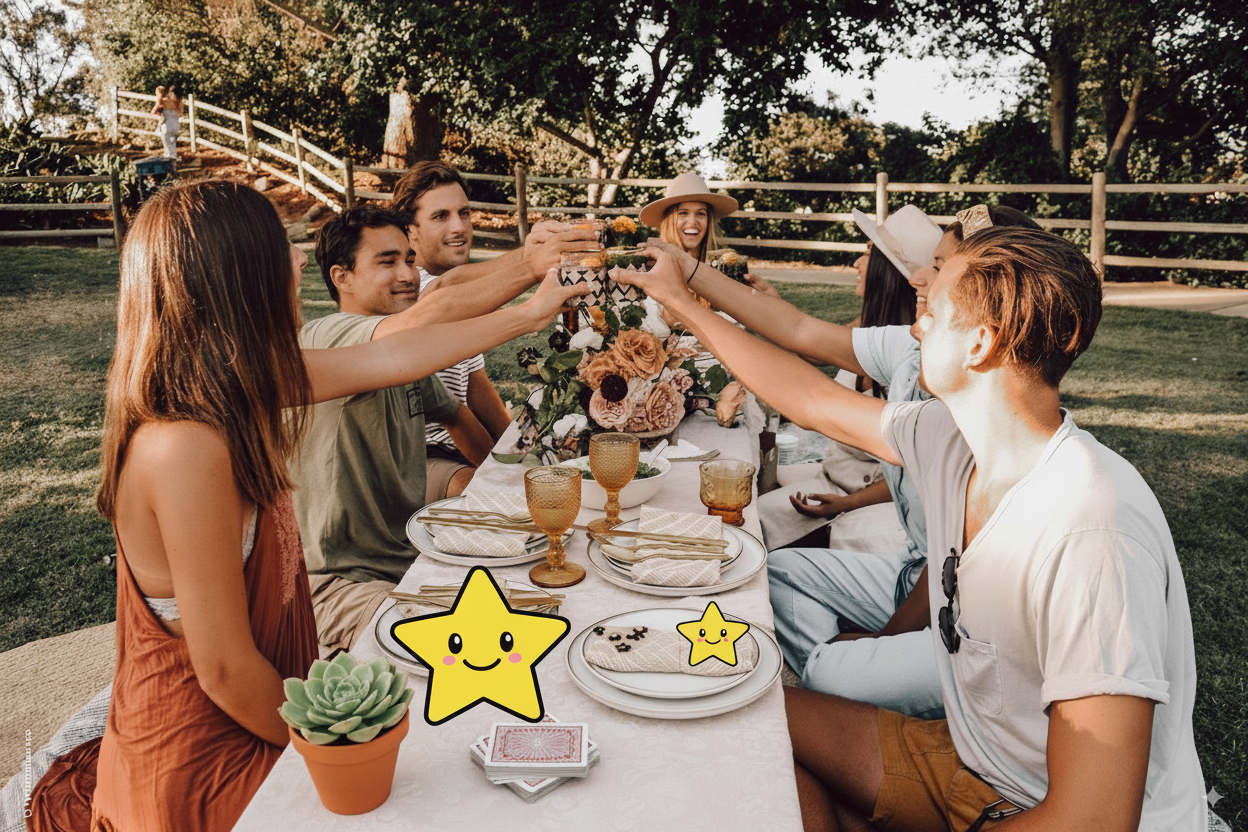}
        \caption{User-defined replacement}
        \label{fig:replacement}
    \end{subfigure}
    \vspace{0.3cm}
    \begin{subfigure}[b]{0.32\textwidth}
        \includegraphics[width=\textwidth]{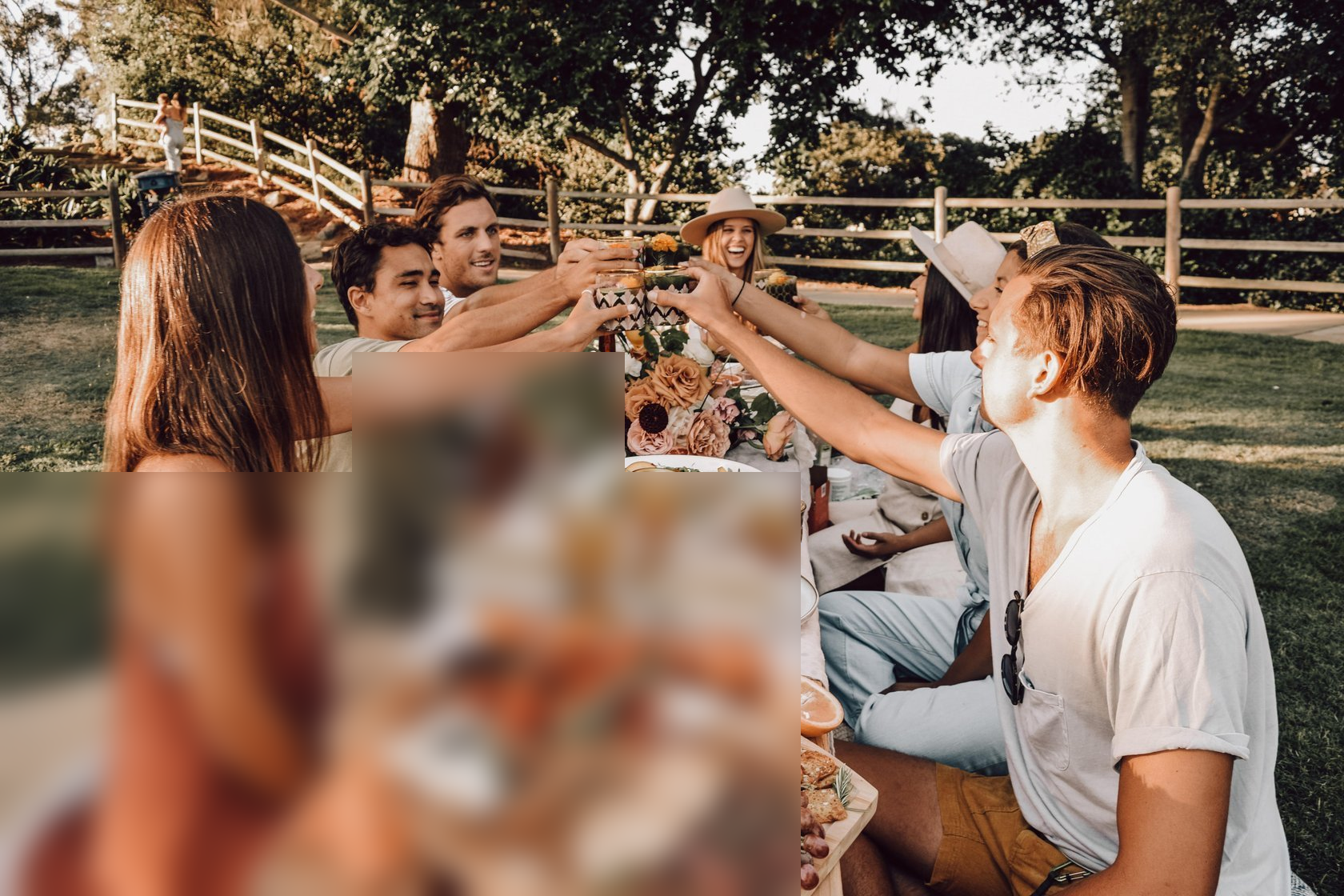}
        \caption{Obfuscation (blur)}
        \label{fig:blur}
    \end{subfigure}
    \hfill
    \begin{subfigure}[b]{0.32\textwidth}
        \includegraphics[width=\textwidth]{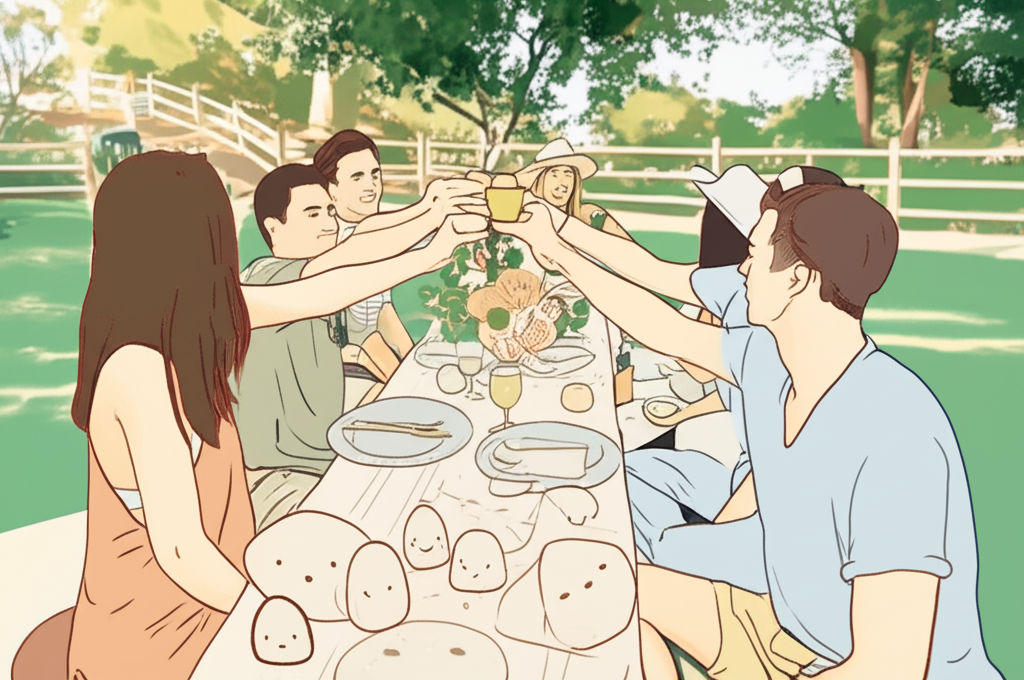}
        \caption{Ghibli Style (Full Image)}
        \label{fig:ghibli}
    \end{subfigure}    
    \hfill
    \begin{subfigure}[b]{0.32\textwidth}
        \includegraphics[width=\textwidth]{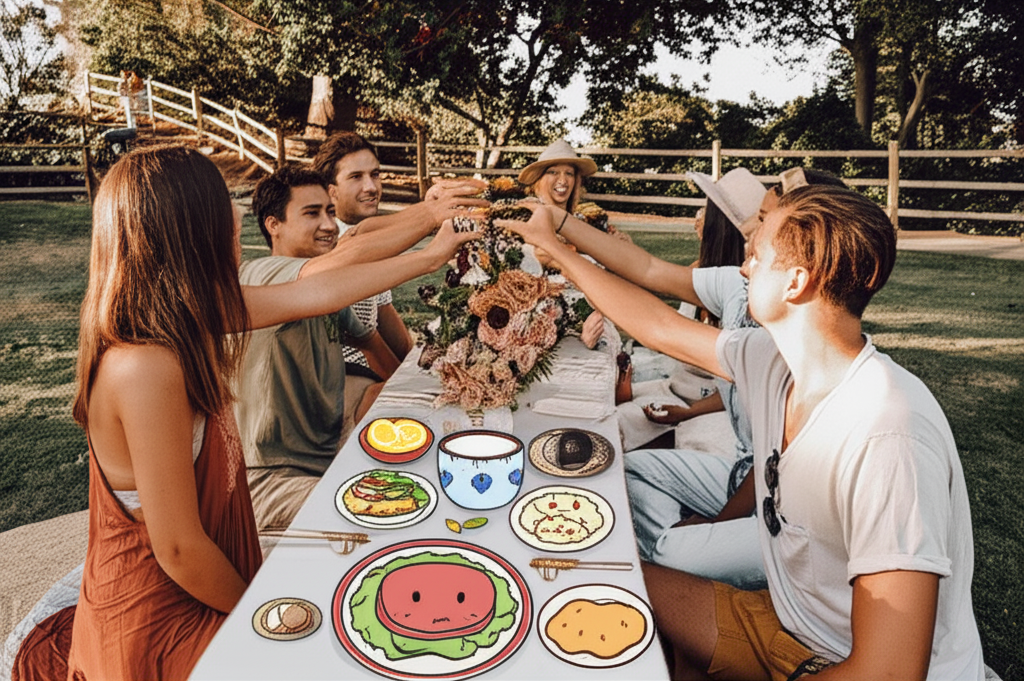}
        \caption{Ghibli Style (Selective)}
        \label{fig:selective-ghibli}
    \end{subfigure}
    
    \caption{Intervention palette demonstrating DIY-MOD's three categories of transformation, applied to an original image (a). \textbf{Semantic Modification[b, c]} alters the content through techniques like (b) inpainting, which removes the trigger and (c) replacement with user-specified alternatives---trees, stars, cards in this case. \textbf{Obfuscation[d]} reduces trigger fidelity, shown here with blurring (d). \textbf{Stylistic Alteration[e, f]} changes the rendering style; a Studio Ghibli style animation is shown applied to the entire image (e) and selectively to only the trigger region (f). The transformations shown address the filter created by Rex (\Cref{sec:intro}), who is managing binge-eating disorder.}
    \label{fig:intervention-palette} 
\end{figure*}

\subsubsection*{Transparency in Modified Content}\label{transparency_modification_indicator}
\diymod{} client marks all modified content with visible cues, displaying a ``Modified by DIY-MOD'' badge at the top right (\Cref{fig:teaser_user_pov}d). When sensitive content dominates a post---making partial modification insufficient---the system displays personalized warning overlays that prevent unexpected exposure. Users can click through to view the original content if they so choose. The visible indicators and override options build trust by giving users control \textbf{(DG2)}.

\subsection{Server Architecture and Processing Pipeline}

The \diymod{} server handles both filter creation and content processing. For filter creation, it maintains stateful sessions with client chat interface. For content processing, it analyzes batches of posts from client adapters.
When users browse supported platforms, the client adapter extracts content from API responses each containing a batch of posts (typically 25 for Reddit) and sends the content to the server. The server uses LLMs and LVLMs to evaluate the text and images in each post against the user's active filters. When a match is found, the server initiates its transformation pipeline to apply an ``intervention''. Each intervention represents a different way to modify sensitive content while preserving informational value. 
\Cref{subsec:interventions}
describes the range of interventions we explored, and \Cref{sec:selection_framework} details our two-stage pipeline for selecting the most appropriate one in real-time for each user and post.

The computational demands of this server processing pipeline are substantial. For example,   processing a single Reddit batch can require 50 to 250 sequential external LVLM API calls that must complete within seconds to maintain usability. \new{In our deployment, the critical path for processing an initial 25-post Reddit batch completed in roughly 5–15 seconds, with subsequent batches typically processed before users scrolled to them.} We achieved real-time performance through \textit{asynchronous processing with multi-batch responses}, \textit{content-based caching}, and \textit{predictive prefetching}. We have described these optimizations and additional design decisions in \Cref{caching_and_prefetching}.

\subsection{The Intervention Palette}
\label{subsec:interventions}

\aptLtoX{\begin{table*}[!h]
\centering
\caption{\note{[Moved from Appendix]} The palette of implemented interventions for image content in \diymod{}.}
\begin{tabular}{lll}
\hline
\textbf{Category}                                & \textbf{Intervention} & \textbf{Description \& Relation to Design Space}                                                                                                                                                                     \\
\hline
\textbf{Obfuscation}                             & Blur                  & Applies a Gaussian blur to a target region, reducing its {}trigger fidelity{} by obscuring details. The surrounding content remains unchanged. \\
\cline{2-3}
                                                 & Occlusion             & Renders an opaque rectangle over a target region. This decreases the {}trigger fidelity{} by covering the area but looks jarring resulting in low {}perceptual smoothness{}.                                      \\
\cline{2-3}
                                                 & Warning\newline Overlay       & Combines Occlusion with a textual warning label placed on top of the hidden region. This eliminates {}trigger fidelity{} while providing explicit context to the user.                                  \\
\hline
\textbf{Semantic Modification}                   & Inpainting            & Removes the regions related to the user sensitivity and reconstructs the background. The result has high {}perceptual smoothness{} at the cost of {}semantic fidelity{}, as it presents a modified version of reality.   \\
\cline{2-3}
                                                 & Replacement           & Replaces a objects related to user sensitivities with a benign one. The user can customize what to replace with. This alters the image's {}semantic fidelity{} but aims for a visually coherent result.                              \\
\cline{2-3}
                                                 & Shrink                & Reduces the scale of the triggering object/region within the image. While preserving the object's presence, thus results in both higher {}semantic fidelity{} and trigger fidelity than inpainting.            \\
\hline
\textbf{Stylistic Alteration\textsuperscript{*}}\newline \footnotesize\textit{---\newline Can apply to \textbf{both} sensitive region only or whole image} & Cubism                & Reimagines the target region through geometric planes and fragmented forms. This high level of abstraction reduces photorealism, creating conceptual distance. \\
\cline{2-3}
                                                 & Ghibli                & Renders a region with the soft, rounded aesthetic of Studio Ghibli animation, reducing photorealism to lower {}trigger fidelity{}.
\\
\cline{2-3}                                                                               
                                                 & Impressionism         & Applies soft, visible brushstrokes and diffused light, creating a dream-like quality that distances the viewer from the literal depiction.
                                                                           \\
\cline{2-3}
                                                 & Pointillism           & Reconstructs a region using small, distinct dots of color, lowering the detail and {}trigger fidelity{} of the original.\\
\hline                                                                                    
\end{tabular}
\label{table:image_interventions}
\end{table*}}{\begin{table*}[!h]
\centering
\caption{\note{[Moved from Appendix]} The palette of implemented interventions for image content in \diymod{}.}
\note{\rule{\linewidth}{2pt}} 
\label{table:image_interventions}
\begin{tblr}{
  width = \linewidth,
  colspec = {Q[135]Q[125]Q[680]},
  cell{2}{1} = {r=3}{},
  cell{5}{1} = {r=3}{},
  cell{8}{1} = {r=4}{},
  hline{1,12} = {-}{0.08em},
  hline{2,5,8} = {-}{0.05em},
  hline{3-4,6-7,9-11} = {2}{l},
  hline{3-4,6-7,9-11} = {3}{r},
}
\textbf{Category}                                & \textbf{Intervention} & \textbf{Description \& Relation to Design Space}                                                                                                                                                                     \\
\textbf{Obfuscation}                             & Blur                  & Applies a Gaussian blur to a target region, reducing its {}trigger fidelity{} by obscuring details. The surrounding content remains unchanged. \\
                                                 & Occlusion             & Renders an opaque rectangle over a target region. This decreases the {}trigger fidelity{} by covering the area but looks jarring resulting in low {}perceptual smoothness{}.                                      \\
                                                 & Warning\newline Overlay       & Combines Occlusion with a textual warning label placed on top of the hidden region. This eliminates {}trigger fidelity{} while providing explicit context to the user.                                  \\
\textbf{Semantic Modification}                   & Inpainting            & Removes the regions related to the user sensitivity and reconstructs the background. The result has high {}perceptual smoothness{} at the cost of {}semantic fidelity{}, as it presents a modified version of reality.   \\
                                                 & Replacement           & Replaces a objects related to user sensitivities with a benign one. The user can customize what to replace with. This alters the image's {}semantic fidelity{} but aims for a visually coherent result.                              \\
                                                 & Shrink                & Reduces the scale of the triggering object/region within the image. While preserving the object's presence, thus results in both higher {}semantic fidelity{} and trigger fidelity than inpainting.            \\
\textbf{Stylistic Alteration\textsuperscript{*}}\newline \footnotesize\textit{---\newline Can apply to \textbf{both} sensitive region only or whole image} & Cubism                & Reimagines the target region through geometric planes and fragmented forms. This high level of abstraction reduces photorealism, creating conceptual distance. \\
                                                 & Ghibli                & Renders a region with the soft, rounded aesthetic of Studio Ghibli animation, reducing photorealism to lower {}trigger fidelity{}.                                                                               \\
                                                 & Impressionism         & Applies soft, visible brushstrokes and diffused light, creating a dream-like quality that distances the viewer from the literal depiction.                                                                           \\
                                                 & Pointillism           & Reconstructs a region using small, distinct dots of color, lowering the detail and {}trigger fidelity{} of the original.                                                                                     
\end{tblr}
\note{\rule{\linewidth}{2pt}} 
\end{table*}}

The design of our system is based on a shift from content \emph{suppression} to \emph{transformation}.  In taking this approach, we faced a design challenge: every intervention has to navigate a trade-off between three competing goals. The first is \emph{semantic fidelity}, or how faithful the modified content is to the original's intended meaning. The second is \emph{trigger fidelity}, or how close the modified distressing element is to its original form. The final goal is
\emph{perceptual smoothness}, or how natural the result of the intervention looks to the user. We designed a palette of transformations, each reflecting a different approach to these trade-offs. \new{Across this palette, interventions range from softening distressing elements to removing them, so people can still engage with the valuable parts of a content without being exposed to their triggers at full intensity. This approach echoes exposure therapy’s emphasis on working with weakened versions of stimuli.}

This palette includes transformations for both text and images. For images, we implemented and explored \textbf{three} categories of transformation: \textbf{obfuscation}, \textbf{semantic modification}, and \textbf{stylistic alteration}. First, \textbf{obfuscation} techniques prioritize the immediate reduction of trigger fidelity. Occlusion (a type of obfuscation) for instance, draws a solid dark rectangle over the region(s) that matches user's sensitivity. This approach aims to preserve the semantic fidelity of the surrounding context by leaving the rest of the image untouched. It scores low on perceptual smoothness, as the edit will be jarring.

In contrast, \textbf{semantic modifications} prioritize perceptual smoothness. Inpainting\cite{yu2018inpainting, li2024mat} as an example case, can seamlessly remove an object and reconstruct the background. This is a greater sacrifice of semantic fidelity than occlusion because it actively creates a new, plausible reality where the object never existed, rather than simply occluding part of the original. Another semantic intervention, \emph{visual euphemism}, explores a different trade-off. It replaces a triggering object with a benign alternative, for instance, substituting spider with a leaf. While this also sacrifices semantic fidelity, the goal is to produce a visually coherent image that we hypothesize is less distressing to the user. To personalize this, users can add additional metadata to their filters or give examples to guide the visual euphemism choices, which may in turn alter the mood of the post.

Third, we also explore how \textbf{altering an image's  rendering style} could offer a middle ground. For example, reducing photorealism with {artistic alteration}, such as ~\emph{Pointillism}~\cite{clement1999neo} can lower the trigger fidelity of a sensitive region while preserving more structural information than occlusion. 
For these alterations, we borrow the aesthetic language of well-known artistic styles---\emph{Pointillism}, \emph{Cubism}, \emph{Impressionism}, and \emph{Studio Ghibli style animation}. We intended for each artistic style to not only reduce realism, but also act as an affective vessel that shapes how content is experienced. Impressionism, with its soft brushstrokes and diffused light, often evokes a sense of tranquility, distance, or dream-like calm. Studio Ghibli inspired style, with its rounded and whimsical rendering, lends even serious scenes a tone of cuteness and warmth. Cubism, which fragments forms in abstract planes, introduces a kind of conceptual distance, perhaps inviting interpretation over emotional reaction.

Similar trade-offs as discussed above apply to text, where interventions range from rewriting texts (prioritizing smoothness) to blurring specific phrases/words (prioritizing fidelity).

\aptLtoX{\begin{table*}[!t]
\centering
\caption{\note{[Moved from Appendix]} The palette of implemented interventions for text content in \diymod{}.}
\label{table:text_interventions}
\note{\rule{\linewidth}{2pt}} 
\begin{tabular}{ll}
\textbf{Intervention} & \textbf{Description \& Relation to Design Space}                                                                                                                                                                                                  \\
\textbf{Blurring }             & Obscures specific words or phrases related to the user sensitivity.                            \\
\textbf{Rewrite   }            & Rephrases a sentence or passage to remove concepts related to the user sensitivity while preserving the original meaning \\
\textbf{Overlay\newline Warning}       & Hides an entire text block behind a personalized warning message.                                  
\end{tabular}
\end{table*}}{\begin{table*}[!t]
\centering
\caption{\note{[Moved from Appendix]} The palette of implemented interventions for text content in \diymod{}.}
\label{table:text_interventions}
\note{\rule{\linewidth}{2pt}} 
\begin{tblr}{
  width = \linewidth,
  colspec = {Q[135]Q[848]},
  hline{1,5} = {-}{0.08em},
  hline{2,3,4} = {-}{0.05em},
}
\textbf{Intervention} & \textbf{Description \& Relation to Design Space}                                                                                                                                                                                                  \\
\textbf{Blurring }             & Obscures specific words or phrases related to the user sensitivity.                            \\
\textbf{Rewrite   }            & Rephrases a sentence or passage to remove concepts related to the user sensitivity while preserving the original meaning \\
\textbf{Overlay\newline Warning}       & Hides an entire text block behind a personalized warning message.                                  
\end{tblr}
\note{\rule{\linewidth}{2pt}} 
\end{table*}}

In \diymod{}, we implemented a representative set of interventions that explore the broader design space of content transformation. \Cref{fig:intervention-palette} illustrates several examples from our image palette's three main categories.  We list all implemented interventions in \new{\Cref{table:image_interventions} and \ref{table:text_interventions}} \remove{\Cref{appendix:intervention_description}}. The range of transformation options presents an important question: \textit{which intervention is most appropriate for a specific user and post?} Next, we detail the framework we developed to address this question.



\subsection{Intervention Selection Framework}
\label{sec:selection_framework}
\label{section:framework}
\begin{figure*}[!t]
    \centering
    \includegraphics[width=\textwidth]{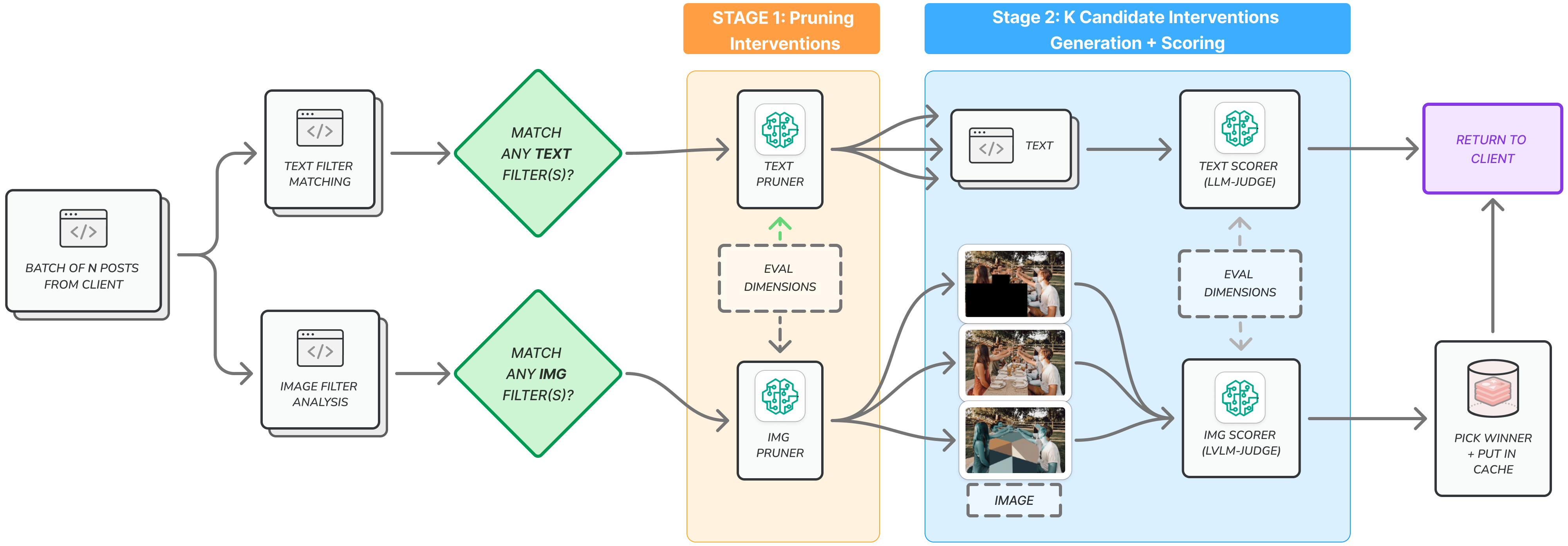}
    \caption{DIY-MOD's two-stage intervention selection pipeline. Content matching user filters enters Stage 1 (pruning) to identify promising interventions, then Stage 2 generates and scores K candidates before selecting the best transformation. Non-matching content bypasses the pipeline entirely (paths not shown for clarity).}
    \label{fig:server_pipeline_figma}
\end{figure*}

We developed a two-stage pipeline (\Cref{fig:server_pipeline_figma}) that evaluates interventions through the lens of individual user needs while maintaining real-time performance. Content that matches a user's filters enters Stage 1 for pruning, and Stage 2
then generates and scores candidate transformations before selecting the best one.

\subsubsection{Evaluation Criteria:}\label{subsec:eval_criteria}

The three dimensions of semantic fidelity, trigger fidelity, and perceptual smoothness characterize the design space, but do not, on their own, determine which intervention is best suited for a given scenario. That choice requires translating these properties into outcomes relevant to a specific user's safety and context. 
To guide this selection and evaluate outcomes, we operationalize a framework that decomposes intervention quality into \textbf{\textit{four}} complementary dimensions, assessed through LLM-as-a-judge~\cite{zheng2023judging, liu2023gpteval}:


\begin{enumerate}
    \item \new{\textbf{Transformation Seamlessness:} This dimension evaluates the perceptual smoothness of the output or the coherence of text. It maps directly to the \textit{perceptual smoothness} axis of our design space, ensuring the intervention is not jarring.}
    
    \item \new{\textbf{Semantic Fidelity:} Inherited directly from our design space, this assesses whether original facts and context are preserved without adding misinformation or hallucinations, a known risk in generative models~\cite{huang2025survey, gao2024aigcs}.}
    
    \item \new{\textbf{Predicted Emotional Safety:} For each user–content pair, this dimension acts as a personalized proxy for \textit{trigger fidelity}. It evaluates how well a given transformation, in light of that user's specific context (filter description, sensitivity level, chat history, and metadata preferences), is expected to reduce trigger exposure and meet the user's safety needs.}
    
    \item \new{\textbf{Contextual Harm Risk:} This dimension functions as a safeguard, assessing the degree to which a transformation might trivialize or distort themes that individuals or broader culture consider sensitive. For instance, applying Studio Ghibli style to war imagery or racial violence would produce a high contextual-harm score. This risk score is subtracted from the overall intervention rating, so transformations that produce contextually inappropriate content are strongly penalized even if they score well on other dimensions.} 
\end{enumerate}

\subsubsection{Two-Stage Selection Pipeline}
Generating all possible interventions for each post to determine the most appropriate one would be prohibitively expensive.
We designed a cascade architecture to address this by reducing computational resource utilization by up to 75\% through predictive pruning, making our selection pipeline practical to use in real-time.

\noindent\textbf{{Stage 1: Predictive Pruning. }}The Pruner \new{uses \texttt{GPT-4o}} to perform\remove{s} multi-task inference\cite{son2024multi} that first checks whether content matches any user filters. For matching content, it predicts which interventions would most likely succeed for that specific user. We provide the Pruner LVLM with detailed descriptions of each intervention including trade-offs, advantages, disadvantages, and ideal use cases. We also supply our four scoring dimensions and instruct the Pruner to evaluate how each intervention would score. This process approximates the full scoring process without generating actual transformations and produces a ranked list from which we select the top-$K$ candidates (typically $K=3$).

\noindent\textbf{{Stage 2: Generation and Scoring. }}
\diymod{} applies the top-$K$ candidates to the content, in parallel. Obfuscation techniques are applied locally using \new{\texttt{GroundingDino}}~\cite{liu2023grounding} and Python Pillow~\cite{clark2015pillow} library. For semantic modifications and stylistic alterations, we use \new{Gemini}'s asynchronous image generation model: \new{\texttt{gemini-2.0-\allowbreak flash-\allowbreak preview-\allowbreak image-\allowbreak generation}}~\cite{google_gemini2_0flash_2025}.
The Scorer \new{uses \texttt{GPT-4o\allowbreak -2024\allowbreak -08-06} to} evaluate\remove{s} each candidate transformation against the evaluation dimensions (\Cref{subsec:eval_criteria}). \new{To mitigate potential biases inherent in LLM-as-a-judge systems where a model may favor its own outputs, we deliberately employ a different model for the scoring phase (\texttt{GPT-4o}) than for the generation phase (\texttt{Gemini 2.0}).} We implement two guardrails in the scoring process. First, the Scorer receives both the original and transformed content, and is instructed to penalize misapplied interventions and failures. Second, we employ two-stage prompting~\cite{wang2024rl} that improve judge-VLM's performance and results in more reliable and consistent evaluations. The Scorer first performs chain-of-thought analysis along our evaluation dimensions given the original content, transformed candidate, and user context. Based on this free-form response, it then produces a final numerical score. These scoring operations run in parallel across all $K$ candidates\new{, and the highest-scoring intervention is selected.}

\remove{The Scorer personalizes evaluation using three user contexts: the filter description, the self-assigned sensitivity level (1-5), and the complete chat history from filter creation. The scores are summed and the highest-scoring intervention is selected as the winner.}


\subsubsection*{Deployment Considerations}
For text transformations and simple image obfuscations (e.g., Occlusion), the system sends lightweight instructions that are applied directly in the user's browser using CSS. For other transformations, the generated content is served to the client. In all cases, we apply modification indicator badge (\Cref{transparency_modification_indicator}) for transparency and store the resulting content in server cache (\Cref{appendix:caching}).


\section{Study 1: In-Situ Evaluation of DIY-MOD}
\label{sec:methods-2-evaluation}
\label{sec:user-study-1}

We evaluated \diymod{} through two complementary studies. This first study examines the system's effect on user agency and safety during naturalistic browsing, evaluating the viability and user experience of personalized content transformation in real-world settings. Where we present participants' free-text responses for both of users studies, we identify them with a string of the form ``P\_1\_'' + a participant identifier.


\subsection{Methods}
\noindent\textbf{Participants:}
We recruited 15 participants through university mailing lists and online communities (9 female, 6 male; ages 21--41). Eligibility criteria included: (1) self-reported sensitivity to specific categories of online content that affect their emotional wellbeing, (2) regular use of Reddit, and (3) being 18 years of age or older. Three participants had also participated in our formative study. Participants received \$20 compensation for completing this study.

\vspace{4pt}
\noindent\textbf{Procedure:}
Participants installed the browser extension and shared their screen via Zoom. Using think-aloud protocol, they described their reasoning while creating filters through the conversational interface. They then browsed their Reddit homepage and subscribed subreddits with \diymod{} active, and narrated their reactions to interventions as they encountered them. We concluded with a semi-structured interview about their setup and usage experience, interpretation of different intervention types, and sense of agency over content exposure. The study lasted approximately one hour for each participant.

\vspace{4pt}
\noindent\textbf{Data Collection and Analysis:}
We transcribed all think-aloud data and interviews, then analyzed them using reflexive thematic analysis~\cite{braun2019reflecting} to identify patterns in user experiences and system interactions. 
During consent, we explicitly warned participants they might encounter content matching their stated sensitivities and emphasized they could pause, skip content, or withdraw at any point.


\subsection{Findings}

The filters created by participants covered a wide range of topics. Some filters addressed the types of acute, personal sensitivities that motivated this work, such as phobias and trauma-related content. Others were created for more common online experiences, such as avoiding spoilers for a popular TV series, discussions about student loans, or even ``content featuring child influencers presented in a clickbaity way''.


\vspace{-8pt}
\subsubsection{Articulating Needs to a Conversational AI}
The study began by exploring whether users could successfully articulate their nuanced safety needs to our system. We found the conversational interface helped users refine broad sensitivities into specific filters. Participants consistently valued \diymod{}'s ability to ask for clarification. This dialogue made defining broad sensitivities, like ``war'', feel more manageable and built confidence that their needs were being understood. As P\_1\_06 noted, this was helpful because \textit{``maybe someone wants to see the news, but they don't want to see the people's discussions and takes on it.''} This sense of being understood was deepened when the system identified abstract relationships between filters. For instance, after P\_1\_08 set a filter for ``antisemitism'', and later added one for ``xenophobia'', the system prompted them to consider whether the new filter should revise the existing one, demonstrating an ability to connect related concepts. While some users were initially unsure how much detail to provide, they adapted quickly. We found the overall ease of use was a consistent theme.

\subsubsection{A Newfound Sense of Agency and Control:}
Participants consistently described feeling empowered by \diymod{}. The tool shifted their stance from reactive defense to proactive control. P\_1\_11 felt: \textit{``more control over the content that is in my feed.''} P\_1\_09 described how \textit{``the ball now is in my court.''}

This control manifested through system affordances. Participants experimented with the sensitivity slider, discovering higher values produced more aggressive interventions. They tested different levels, verified the system's response, and calibrated settings to match their emotional need tied to the filter. Users further customized filters through the options page: attaching metadata to guide intervention selection, specifying preferences for certain kinds of interventions (e.g., warnings over rewrites) or defining which benign objects should replace triggers in visual euphemisms.

Some participants feared creating what they half-jokingly called their own ``echo chamber'' or ``bubble''. To these participants, occasional system flaws in detection of unsafe regions or semantic replacements were in fact useful disruptions. 
P\_1\_13 articulated why the occasional failure mattered: \textit{``I don't feel that it's psychologically healthy to feel like you have 100\% control, because that's not how the real world works.''} Consistent with this view, participants valued the ability to click through the post to reveal unmodified content when the intervention \textit{``was too heavy-handed.''} P\_1\_07 noted the tool was effective at \textit{``preventing me from seeing it if I'm just casually scrolling without completely blocking access.''}

\diymod{} provided a way to manage personal preferences without the social friction of platform reporting, alleviating a tension several participants described. Users explained that reporting content that is personally distressing but does not violate community guidelines can feel both ineffective and unfair to the content creator.
Several participants framed using the tool as an act of  ``self-care'', allowing them to manage their own well-being without publicly penalizing others.

\subsubsection{Trust is Built on Transparency.} Trust in the system was overwhelmingly tied to its transparency. The visual indicators that marked modified content were universally seen as essential for knowing when the system was acting on their behalf and for distinguishing reality from modification. P\_1\_11 explained: \textit{``The indicators were essential. They built trust and, even when things failed, they helped me understand that this was related to my filters and sensitivities.''} 

\subsubsection{Views on Sharing Filter Configuration:} We wanted to know whether there would be value in a social ecosystem where filter configurations could be shared, reused, and adapted across users to support one another's wellbeing. Participants saw value in sharing filters, but only within trusted relationships. An immediate use case they identified was for {intergenerational care}. P\_1\_01 expressed excitement about sharing her filters she setup during session 1 with one of her parents who has the same phobia as her. To her, this would be a perfect and immediate way to provide care for a loved one. This idea of filters as a tool to protect children or assist less tech-savvy relatives was a recurring sentiment (P\_1\_02, P\_1\_03, P\_1\_06).  Outside that trusted circle, some participants worried about judgment and privacy. P\_1\_10 feared that \textit{``people would judge me based on what filters I set''}, while P\_1\_02 simply found their filters ``too personal'' to share with others. P\_1\_07 wanted to selectively share their filters with a particular community without disclosing more personal ones.

\subsubsection{Limits of Naturalistic Exposure:}
During the one-hour study session, some participants did not organically encounter the kinds of posts that matched their filters. Some of them decided to stress-test \diymod{} by visiting subreddits they knew were likely to contain sensitive content. Before they did so, we reminded them again that sometimes interventions could fail and confirmed they were comfortable with proceeding.
This naturalistic browsing also led to imbalanced exposure to some intervention types. For example, for users with text-based filters, we noted \diymod{} frequently relied on personalized overlays. This was because other transformations were often unsuitable. A ``rewrite'' could not preserve the meaning if an entire post was about a sensitive topic, and blurring the whole text would eliminate all informational value. 


This imbalanced exposure combined with the ethical constraints of asking participants to actively seek potentially distressing content, motivated our second study. To understand the nuanced factors that drive user preferences between different transformations and how to improve our intervention selection framework, we needed a more controlled setting which allowed for systematic exposure to a broad range of intervention types.

\begin{figure*}[!t]
    \centering
    \includegraphics[width=0.6\linewidth]{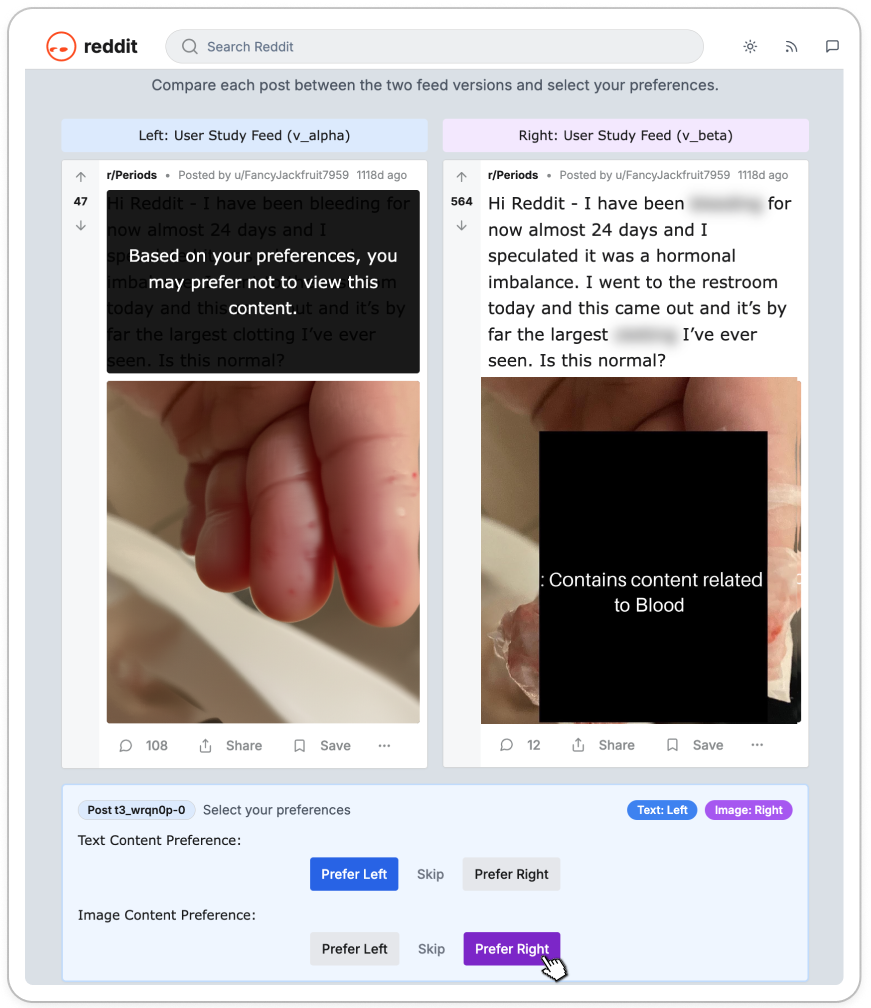}
    \caption{\note{[Moved from Appendix]} The custom application used in Study 2 presented participants with a side-by-side, synchronized dual feed. Each post appeared as a pair of distinct modifications. Participants scrolled the feed and, for each post, used the controls at the bottom to indicate their preference for the text and image transformations separately. In this example, the user has preferred the text transformation on the left (a full overlay) but preferred the image transformation on the right (occlusion). Modification indicators were not shown in this interface, as participants were informed that all content was transformed by the system.
}
    \label{fig:dual-feed-sample}
    \vspace{-15pt}
\end{figure*}

\section{Study 2: Understanding User Preferences for Content Transformation}

\label{sec:user-study-2}
\label{sec:user_study_finding}

This study identifies the principles behind effective content transformations. To achieve this, we presented participants with pairs of modified content and analyzed their preferences and rationales to extract these underlying principles. This analysis provides a deeper understanding of user needs and also yields a dataset of pairwise preferences to inform the development of more user-aligned selection models.

\subsection{Method} 
\noindent\textbf{Materials and Procedure:}
We developed a custom \texttt{Next.js} application with a dual-feed interface that presented two parallel, synchronized versions of a feed side-by-side replicating Reddit's UI. For each participant, we curated a personalized feed of 10–15 posts relevant to the sensitivities they had configured with a permanent duration in Study 1. All posts were pre-screened to ensure they were representative of the participant's sensitivities while excluding gratuitously graphic or extreme material.

As participants scrolled through the dual feed, they encountered two distinct, modified versions of each post with text and image transformation applied (\remove{details in ~\Cref{appendix:dual_feed}}\new{\Cref{fig:dual-feed-sample}}). For each pair, they were asked to select the version that better met their safety needs and explain their reasoning. The two modifications were chosen systematically: one was the intervention ranked highest by \diymod{}'s selection framework, while the other was randomly selected from the framework's other high-ranking alternatives (ranks 2-5). To mitigate positional bias, the on-screen placement (left or right) of these two versions was randomized for each post.
While the primary task involved comparing the two modified versions, the original unmodified content was also available for context. Participants could request to view it or have it described verbally if needed. The study lasted approximately 30 minutes for each participant.

\vspace{4pt}
\noindent\textbf{Participants:} We invited all 15 participants from Study 1 to this study. Twelve participants completed study 2 (7 female, 5 male; ages 23–41). Each received \$10 in compensation.

\vspace{4pt}
\noindent\textbf{Analysis: }
We analyzed participants' qualitative justifications using thematic analysis to understand the contextual reasons for preferences. For the quantitative analysis of how well the system's selected interventions matched our participants' preferences, we modeled the probability of a preference match using generalized linear mixed-effects models with participants as random effects.

\subsection{Findings}
\subsubsection{Strategies for Achieving Safety Through Transformation}
Participants selected transformation strategies based on the perceived informational value of content and the severity of their personal trigger.
For content with informational or social value, participants preferred interventions that eliminated distressing elements while maintaining context. P\_1\_08, viewing a vaccine infographic with a baby receiving an injection, preferred inpainting that removed the needle: \textit{``from the position of the hand and the arm, I can understand that a syringe is probably there... I'm glad that this is gone with other things untouched.''} Similarly, P\_1\_09 preferred artistic rendering over blurring for a post about veteran self-harm because it \textit{``gave him enough context to get an idea on what's going on''} while abstracting distressing details.

For content with low perceived informational value, users prioritized creating perceptual distance from the trigger. For instance, P\_1\_05, viewing a post about a political conflict he wished to avoid, preferred an abstract cubist painting intervention. This priority was also evident when a trigger was severe, regardless of the content's perceived informational value. P\_1\_10, who experiences severe distress from crime-related imagery, was shown a photo post on the topic. She immediately chose an occlusion that completely blocked the scene over a context-preserving blur: \textit{``100\% the left one. I'm not seeing any part of this.''}

The same participant made different trade-offs depending on context. P\_1\_12 wanted a random photo of animal harm completely occluded but preferred an impressionist rendering for a news post about ``Illinois Police caught [animal harmer]....'' with a photo of the harmed dog. She had been following this case, so the artistic transformation helped her stay informed on that case reducing graphic details. These varying preferences demonstrate that intervention preference depends on immediate goals of the user for that specific content.

\subsubsection{The Cognitive Impact of Interventions}
An intervention's success is not just about what it visually hides, but about the cognitive state it induces in the user. Successful interventions provide \emph{cognitive closure}, allowing the user to safely disengage. Failed interventions create a \emph{cognitive gap}, an ambiguous state that prolongs mental engagement with the topic the user sought to avoid.

\noindent\textbf{Interventions fail when they create a \emph{gap}.} Blurring or partial removals were often rejected because they created an \emph{information gap} that provoked unhelpful reactions. This gap could result in curiosity, as P\_1\_09 noted that blurred text \textit{``makes me want to know more,''} prompting a desire to look at the original content.  In other cases, the information gap invited users to mentally fill in what was missing. For instance, P\_1\_15 disliked a seamless removal because it made them feel \textit{``like something is being hidden,''} which \textit{``leaves my brain... stuff to my imagination.''} Tonal inconsistency also created a \emph{dissonance gap}. For instance, when shown a stylistic intervention that our system had already rejected due to a poor contextual fit, P\_1\_01 rejected it because the aesthetic clash was simply \textit{``too weird to look at.''}

\noindent\textbf{Interventions succeed when they provide \emph{closure}.}
This closure took two primary forms. The first, \emph{informational closure}, gives users enough context to understand a scene's nature without revealing harmful details, allowing them to avoid further mental engagement. We observed this was achieved through different means. For example, P\_1\_08 preferred an impressionistic rendering that abstracted the harmful details of a self-harm image while preserving the scene's overall context. In a different case, P\_1\_07 found closure through direct explanation, valuing an overlay with a textual warning that stated why an image was hidden. 
Alternatively, interventions achieved \emph{emotional closure} by replacing a trigger with a benign, cognitively complete scene. Rather than leaving an ambiguous void, this provides a complete, non-threatening image that allows the user's mind to immediately disengage, a strategy favored by P\_1\_04. This was particularly effective when the replacement was humorous, as P\_1\_09 noted it can ``take something that might be harmful and make it helpful.''

\subsubsection{Quantitative Alignment with User Preferences}
To complement our qualitative findings, we modeled the probability that participants would select the system's highest-ranked intervention over a nearby alternative---the randomly selected intervention from ranks two through five. 
We use an intercept-only model to predict our dependent variable, \texttt{System Choice Favored}, which is a binary variable equal to 1 if the participant selected the system's choice, and 0 otherwise.
We use the \texttt{glmer} function from 
the \texttt{lme4} package~\cite{lme4} in \texttt{R}.
For image interventions, the estimated intercept was positive and statistically significant (log-odds = 0.53, $p = .033$), corresponding to a 63\% probability that participants chose the system’s top candidate over the close contenders. For text interventions, the alignment was stronger (log-odds = $1.33$, ${p} < .001$), corresponding to a 79.1\% probability.

\section{Discussion}

\subsection{Transformation Enables Access, Not Avoidance}

Empowering users to take moderation decisions into their own hands might raise concerns about whether this agency can trap them in filter bubbles~\cite{pariser2011filter}. \diymod{} can support a different dynamic. By transforming how content appears rather than removing it entirely, the system enables users to engage with content they would otherwise avoid completely. 

Our participants described abandoning entire online communities due to unpredictable triggering content. P\_0\_04 stopped visiting support forums despite needing community connection. P\_0\_12 avoided all social media after pregnancy loss, cutting off social support when she needed it most. These users faced a binary choice: risk psychological harm or disconnect entirely.
\diymod{} creates a third option. Users can engage with challenging topics and diverse viewpoints because the presentation layer that tends to provoke overwhelm or distress has been modified. The information remains accessible while the harm is reduced. 
This sentiment of having a new way to engage with the digital world was articulated by P\_1\_03: 
\begin{quote}
\textit{``I think what you've done is life-impacting for people. Even if it's not perfect, I'm still thankful for it, and I think other people are too.''} 
\end{quote}

At the same time, this approach surfaces deeper tensions: Who decides what users should or should not engage with, and why does that decision need to be made in a top-down, prescriptive manner? The First Amendment is often framed as a protection of speech. Legal scholars have similarly argued for the right to freedom of listening and the right against compelled listening~\cite{corbin2009first}. While these protections do not extend to platforms as private entities, the principles that they represent remain relevant. Our studies show that content that appears benign to most can be experienced as harmful by others. This variability complicates any universal judgment about what content is ``safe'' or ``appropriate''---at least outside the categories codified by law. And it raises a critical question: Who gets to decide which experiences of harm are valid and which users are entitled to protection?


\subsection{The Case for Platform-Integrated Personalization}\label{discussion:case_for_platforms}

While \diymod{} demonstrates personalized moderation is viable as middleware~\cite{fukuyama2021}, its architecture requires intercepting, externally processing, and then scoring generated transformations. These steps introduce unavoidable latency and complicate implementation.

Platforms however, could avoid these bottlenecks entirely. With direct content access, computational resources, and existing user behavioral signals, they could implement personalized moderation far more efficiently. Platforms already use this infrastructure for ad targeting and safety interventions like suicide prevention~\cite{facebook_suicide_1,facebook_suicide_2}. The same systems could also power consensual, user-directed personalized content transformation.

The question becomes one of priorities. \textit{How much more could be achieved if platforms recognized emotional safety as a core user need rather than an afterthought?} We position \diymod{} not as the final solution, but as a starting point that demonstrates what is possible. Platforms should recognize that prioritizing user wellbeing aligns with their own goals. Our study shows that safer users are more engaged users, maintaining connections they would otherwise abandon.

Our approach gives rise to a natural question: Which kinds of content should be addressed through platform-led moderation, and which might be better suited to user-driven transformation? 
Platforms are widely expected to bear responsibility for addressing harms that have broad, collective consequences such as coordinated manipulation campaigns or abuse like hate speech or targeted harassment that fosters hostile environments.
Yet even here, the boundaries are far from settled. The same activity can be cast as harassment or as activism, as disinformation or as political dissent. Platforms' judgments about what to enforce are never neutral. They are shaped by political pressures, business incentives, and cultural norms~\cite{diaz2021double,jahanbakhsh2023thesis}.

Our approach does not resolve these disputes, nor could it. What it does is treat harms as actionable even when there is no consensus on whether content is harmful. It does so without requiring platforms to impose universal judgments that risk overreach, silence expression, and flatten disagreement.

\subsection{Authorship}

Our approach also raises questions about how post authors perceive transformations of their content. Some might welcome transformations if they allow their posts to remain accessible to audiences who might otherwise avoid them. Others may worry that transforming their content, even selectively, changes how their voice or emotional expression is perceived. This reflects a mutual claim to agency in shared spaces: just as the posters have the right to express themselves, viewers have the right to shape how they experience that expression. But these rights can come into tension, even if the substance of the post remains unchanged.
Future work should examine when these transformations are seen as respectful accommodations vs distortions of intent.

\subsection{Application to Civic Discourse}

This transformation-based approach may also hold promise for other types of content not brought up by our study participants. One can imagine its application to civic or ideological discourse: when valuable perspectives are obscured by hostile or inflammatory framing, repackaging that content (e.g., by softening antagonistic language, reducing confrontational cues, or highlighting shared values), could make it easier to engage with. It is also worth noting that there is no clear consensus that exposure to counter-attitudinal information is always beneficial. In fact, a body of research shows that such exposure, especially when unfiltered or confrontational, can exacerbate polarization and deepen animosity~\cite{tornberg2022digital_echo,bail2022breaking}. This calls into question that more exposure is always better, and highlights the need for systems that support constructive engagement, rather than unmediated confrontation~\cite{bail2022breaking}. Altering the tone or phrasing of civic or political content through transformations like ours may offer a path toward openness without overwhelm.
\new{It is also worth considering \textit{who} is most likely to adopt such tools. Prior work suggests challenge-averse individuals may use personalization to reinforce existing views~\cite{munson2010presenting}. However, those findings reflect systems that offer a binary choice (e.g., show/hide). By contrast, transformation reduces the affective load of counter-attitudinal content. It is plausible that this, unlike binary filtering, might enable engagement from users who would otherwise disengage completely. Investigating these adoption patterns remains a critical area for future research.}

Prior work on bridging divides has largely focused on identifying and ranking content that resonates across ideological lines~\cite{ovadya2022bridging, stray2021designing}. While valuable, such approaches depend on the availability of naturally bridgeable posts. But what if we could \emph{make} more content bridgeable? This of course, raises important questions about authenticity and message integrity: How much transformation is too much? When does reframing a message become misrepresentation? Future work must address these ethical boundaries.


\subsection{Therapeutic Grounding and Applications of Our Work}

\diymod{}'s design is inspired by exposure therapy: enabling safer engagement with a weakened or modified version of a stimulus, rather than avoidance~\cite{rothbaum2002exposure,abramowitz2019exposure}. This principle is reflected in our palette of interventions. Our transformations are designed to function as weakened versions of the original content, and the ability to easily adjust sensitivity levels and filter durations echoes the concept of graduated exposure. 

\new{\noindent While \diymod{} is not a clinical intervention in its current form, its combination of weakened content, adjustable sensitivities, and time-bounded filters hints at how similar mechanisms could eventually complement therapeutic practices. For example, these knobs could let users (and clinicians) gradually decrease intervention intensity over time, if they so choose, rather than assuming static sensitivities. This direction aligns with calls~\cite{wang2025mentally} for practitioner-assisted AI systems that support personalized, adaptive exposure in everyday contexts.}

\subsection{Design Considerations of DIY-MOD}

\subsubsection{Architectural Safeguards:} Using AI to interpret personal sensitivities carries inherent risks, as system failures can expose users to content they explicitly sought to avoid. One way we mitigate these risks is by implementing a two-stage cascade architecture: a pruner and then a scorer VLM-judge evaluate content using a consistent rubric. Such multi-stage evaluation strengthens the reliability of LLM/VLM-judge based pipelines~\cite{badshah2024reference, gu2024survey}. \new{We manually audited sampled content and found that all sampled text edits and approximately 89\% of sampled image edits were successfully transformed, with false obfuscations (edits that did not match the viewer’s filter description) in only about 7\% of transformed images (\Cref{appendix:failures}).}

\subsubsection{Privacy by Design.}
Privacy considerations are fundamental when handling data tied to personal sensitivities. \diymod{} adopts a ``privacy-by-design'' approach~\cite{cavoukian2009privacy} that prioritizes user control and data minimization. It provides full functionality without an account and offers user-controlled portability of filters through an anonymous export/import feature. By default, it uses local-first storage and practices data minimization, sending only public in-feed content for ephemeral analysis. Looking ahead, portions of the analysis can migrate to lightweight on-device models to further reduce reliance on server-side processing.

\subsubsection{Collaborative Safety Standards.}\label{disc:collab_safety} The export/import feature enables collaborative safety standards to emerge organically within communities. Our study revealed immediate use cases for filter sharing, from family members with shared phobias to support groups developing collective standards. This approach fundamentally differs from current platform moderation where content rules and digital spaces remain tightly coupled. Today, adopting different moderation standards often means migrating to new spaces. DIY-MOD decouples these layers, allowing users to inhabit shared digital spaces while customizing their individual content experience. However, our findings also highlight a key design challenge for such a sharing ecosystem: balancing the desire for community care with the need for personal privacy.

\section{Limitations and Future Directions}
Our work has several limitations that offer avenues for future research. 

\noindent\textit{Improving Selection Model.}
Our findings reveal that the effectiveness of an intervention is context-dependent, and that successful transformations manage a user's cognitive state by providing closure. Our system's selection framework, in contrast, relied on a limited user model derived from filter descriptions and lacked an explicit model of cognitive closure or the user's immediate context.  Improving its ability to select transformations that provide such closure is an important area for future work. The principles and preference data from this study provide a direct path to do so by aligning the selector with human preferences using alignment approaches such as DPO~\cite{rafailov2023direct}.


\noindent\textit{{Long-Term Psychological Implications.}}
Our study focused on users' immediate experiences, which raises a longitudinal question: do personalized transformations support long-term wellbeing, or can they enable patterns of unhealthy avoidance? There is good reason to shield users from repeated triggers, as re-exposure may exacerbate anxiety~\cite{markowitz2020exposure}. At the same time, indefinite avoidance may also have adverse psychological and social effects. Future work should explore how to design for this tension perhaps by incorporating features that allow users to optionally and safely decrease intervention intensity over time, aligning the tool more closely with therapeutic goals.
 

\noindent\textit{Scope of Evaluation.} 
Our user studies, while providing nuanced qualitative insights \new{through a multi-stage design, involved a formative study with 12 participants, an in-situ evaluation with 15, and a preference elicitation with 12}. This limited scale, along with a primarily Western participant pool, means broader deployment is needed to understand potential cultural variations in how harm is perceived and what transformations are considered appropriate.

\noindent\textit{Technical and Platform Dependencies.} While \diymod{}'s core logic is generalizable, the content interception requires platform-specific adapters, and extending \diymod{} to video-first (e.g., TikTok) platforms would require additional engineering effort. Further, our reliance on commercial VLMs means their capabilities, biases, and safety filters can affect performance. \new{We detail specific failure modes and mitigations in \Cref{appendix:failures}; developing benchmarks to quantify edit failures and over-detection rates remains a potential future work.} 

\new{\noindent\textit{Scalability \& Latency.}    Our middleware approach also introduces computational and monetary costs. Our optimizations (detailed in \Cref{appendix:performance}) keep the \textit{per-batch} latency within a usable range for real-time interaction, but high cost associated with LLM/VLM usage remains a scalability constraint. These constraints highlight the efficiency gains of a platform-integrated approach (\Cref{discussion:case_for_platforms}). Future work could also explore smaller, specialized on-device models to reduce both latency and cost.}\remove{, motivating future work on fine-tuning open-source models.}

\noindent\textit{Future Design Directions.} Participants requested finer-grained collaboration and filtering controls: the ability to export individual filters with signatures for trust, and temporal controls beyond the three presets.

\new{
\noindent\textit{Misuse Potential}.
The same open-ended agency that enables protective filtering could be inverted to highlight content that might cause harm to users themselves. This risk is inherent to any system that provides users with fine-grained control over content presentation. In practice, such misuse is bounded by the safety constraints of the underlying VLM. The system cannot execute interventions that the foundation model itself refuses to generate.}



\section{Ethics and Positionality}
\noindent\textbf{Authors Positionality.} The research is partly motivated by the last author's lived experience with phobias, providing a personal understanding of the challenges of navigating online content with specific sensitivities. 
Throughout the design process, we also consulted with two mental health practitioners specializing in anxiety disorders and exposure therapy to ensure our approach was responsibly grounded.

\vspace{4pt}
\noindent\textbf{Ethical Conduct.}
One key challenge of this work was studying online harm without causing further distress, so our approach was guided by a principle of care. This motivated our two-phase user study protocol. All participants, at all point had the right to skip segments/withdraw without any consequence. Architecturally, it also informed our privacy-by-design choices to protect users' sensitive filter data. All research activities were approved by our institution’s IRB. Throughout this project, we sought to balance the pursuit of knowledge with a respect for the wellbeing of the individuals who made this work possible.

\section{Conclusion}
This work demonstrates a path away from universal, platform-enforced rules towards tools that give individuals agency over their online experience. We argue that status-quo moderation approaches are blunt instruments that eliminate entire posts deemed harmful when the content can in fact be salvaged. We propose an approach that instead modifies the unsafe elements in content while preserving its semantics. We build this new paradigm into a browser extension. We evaluate it through two user studies and demonstrate that users find value in this system which empowers them to remain engaged with communities and content they would otherwise avoid. Ultimately, our approach rethinks moderation not as centralized gatekeeping, rather a personal tool for safely navigating the digital world.

\begin{acks}
We thank Jonathan Stray and Ian Baker for providing access to the codebase for the Prosocial Ranking Challenge. We are grateful to Dr. Elizabeth Duval, Shmeelok Chakraborty, Shwetha Rajaram, Rashidujjaman Rifat, Erika Arias, Rashon Poole, Larnell Moore, and Teanna Sims for their support and feedback throughout the project. Rayhan Rashed also thanks Ehsan Hoque for mentorship that continues to shape how he approaches writing, particularly paper intros. 
\end{acks}

\bibliographystyle{ACM-Reference-Format}
\bibliography{base}
\begin{appendix}

\section{Performance Optimizations for Real-Time Content Processing}
\label{appendix:performance}
\label{caching_and_prefetching}
\subsection{Overview of Computational Load}
Processing personalized content modifications in real-time presents a significant computational challenge. The system must analyze large batches of content and make multiple, time-consuming VLM calls, all within the few seconds a user spends on a single screen of content. To quantify this challenge, we first define the key variables.

Let $N$ be the total number of posts in a batch (typically 25 for Reddit), with $N_t$ being the number of posts containing text and $N_p$ the number containing images. From these, let $M_t$ be the number of posts with text that match a user's filter and $M_p$ be the number of posts with images that match. Finally, let $K$ be the number of intervention candidates our system generates for each matching image (typically $K=3$).

The following table details the different processing sequences, their API call latency, and their dependencies.

\begin{table}[th!]
\centering
\caption{Computational characteristics of the processing pipeline. The pipeline is divided into sequences that must complete on the critical path before a response is sent to the client, and a final sequence that can run asynchronously in the background.}
\label{tab:api_complexity}
\begin{tabular}{lccc}
\toprule
\textbf{Operation} & \textbf{API Calls} & \textbf{Latency} & \textbf{Blocking} \\
\midrule
\multicolumn{4}{l}{\cellcolor{green!10}\textit{Text Processing Sequence (Critical Path):}} \\
\cellcolor{green!10}\quad Filter matching &\cellcolor{green!10} $N$ &\cellcolor{green!10} 0.5-1s &\cellcolor{green!10} Yes \\
\cellcolor{green!10}\quad Intervention selection &\cellcolor{green!10} $M_t$ &\cellcolor{green!10} 1-2s &\cellcolor{green!10} Yes \\
\cellcolor{green!10}\quad Intervention application &\cellcolor{green!10} $M_t$ &\cellcolor{green!10} 1-2s &\cellcolor{green!10} Yes \\
\midrule
\multicolumn{4}{l}{\cellcolor{green!10}\textit{Image Filter Analysis (Critical Path, parallel with text):}} \\
\cellcolor{green!10}\quad Filter matching &\cellcolor{green!10} $N_p$ &\cellcolor{green!10} 2-5s &\cellcolor{green!10} Yes \\
\midrule
\multicolumn{4}{l}{\cellcolor{blue!10}\textit{Image Transformation  (Asynchronous, Background Task):}} \\
\cellcolor{blue!10}\quad Intervention generation &\cellcolor{blue!10} $K \times M_p$ &\cellcolor{blue!10} 2-10s &\cellcolor{blue!10} No \\
\cellcolor{blue!10}\quad Intervention scoring &\cellcolor{blue!10} $K \times M_p$ &\cellcolor{blue!10} 2-5s &\cellcolor{blue!10} No \\
\bottomrule
\end{tabular}
\end{table}

As the table shows, the total number of VLM calls ($T_{\text{total}}$) for a single batch can be expressed as the sum of calls for the text and image pipelines:
\begin{align}
   \text{Text Pipeline:} \quad & T_{\text{text}} = N + 2\times M_t \\
   \text{Image Pipeline:} \quad & T_{\text{image}} = N_p + 2\times KM_p \\
   \text{Total:} \quad & T_{\text{total}} = N + 2\times M_t + N_p + 2\times KM_p
\end{align}

To illustrate the computational load, consider a user with a sensitivity to violent imagery browsing Reddit during a major international conflict. A general-interest subreddit like \texttt{r/worldnews} might have nearly every post contain a matching image. In this case, $M_t \approx N$, $N_p \approx N$, and $M_p \approx N_p$. The total API calls would approach $10N$, or roughly 250 calls for a single 25-post batch, all of which must be processed in seconds to ensure usability.

The table's first two groups, the \textit{Text Processing Sequence} and the \textit{Image Filter Analysis}, represent the critical path that must complete before returning an initial response to the client \new{for the full batch of posts}. The final \textit{Image Transformation Sequence} can then process asynchronously in the background \new{generating time-consuming image transformation and scoring}. Without any optimizations, this pipeline would require several minutes to complete; our approach reduces the critical path latency to 5-15 seconds \new{for the batch}.

\subsection{Asynchronous Processing with Multi-Batch Responses}

\begin{figure*}[t]
  \centering

  \IfFileExists{images/DIY-MOD_swimlane.png}{
    \includegraphics[width=\textwidth]{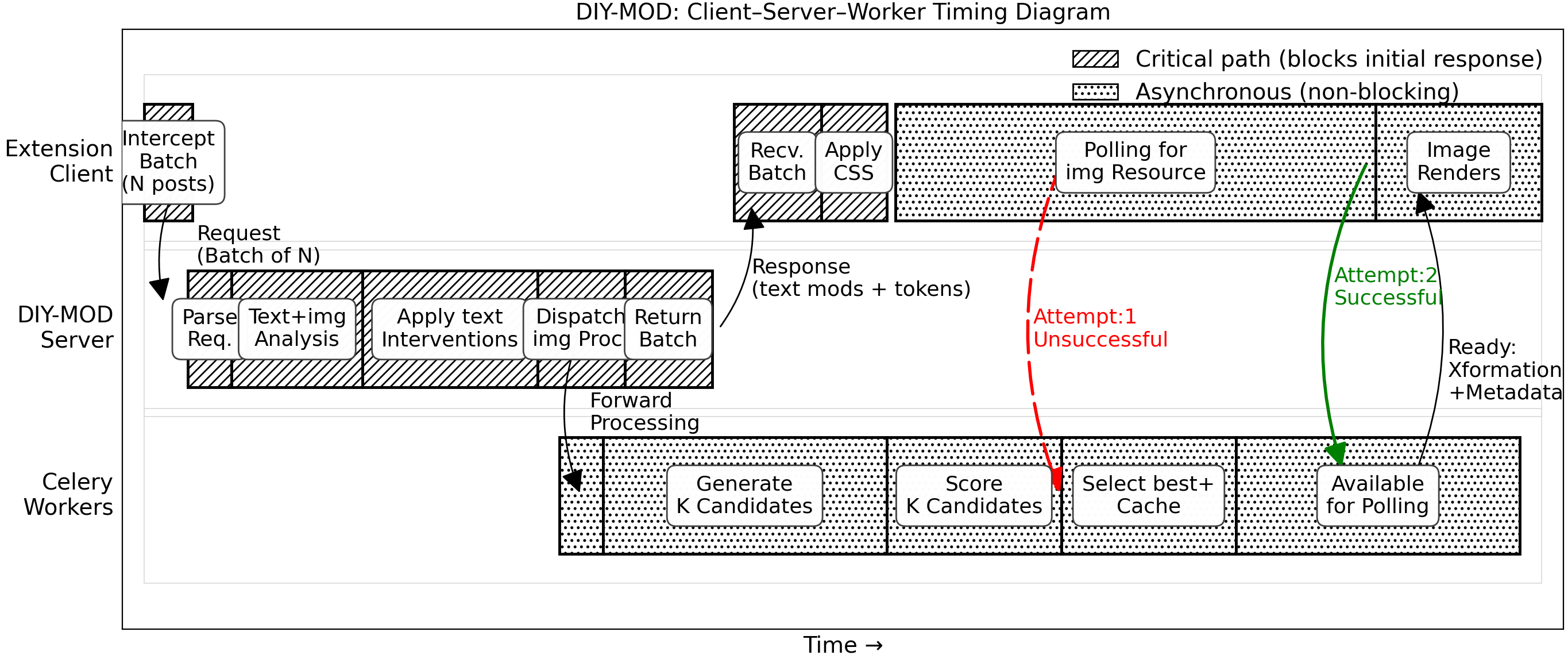}
  }{
    \includegraphics[width=\textwidth]{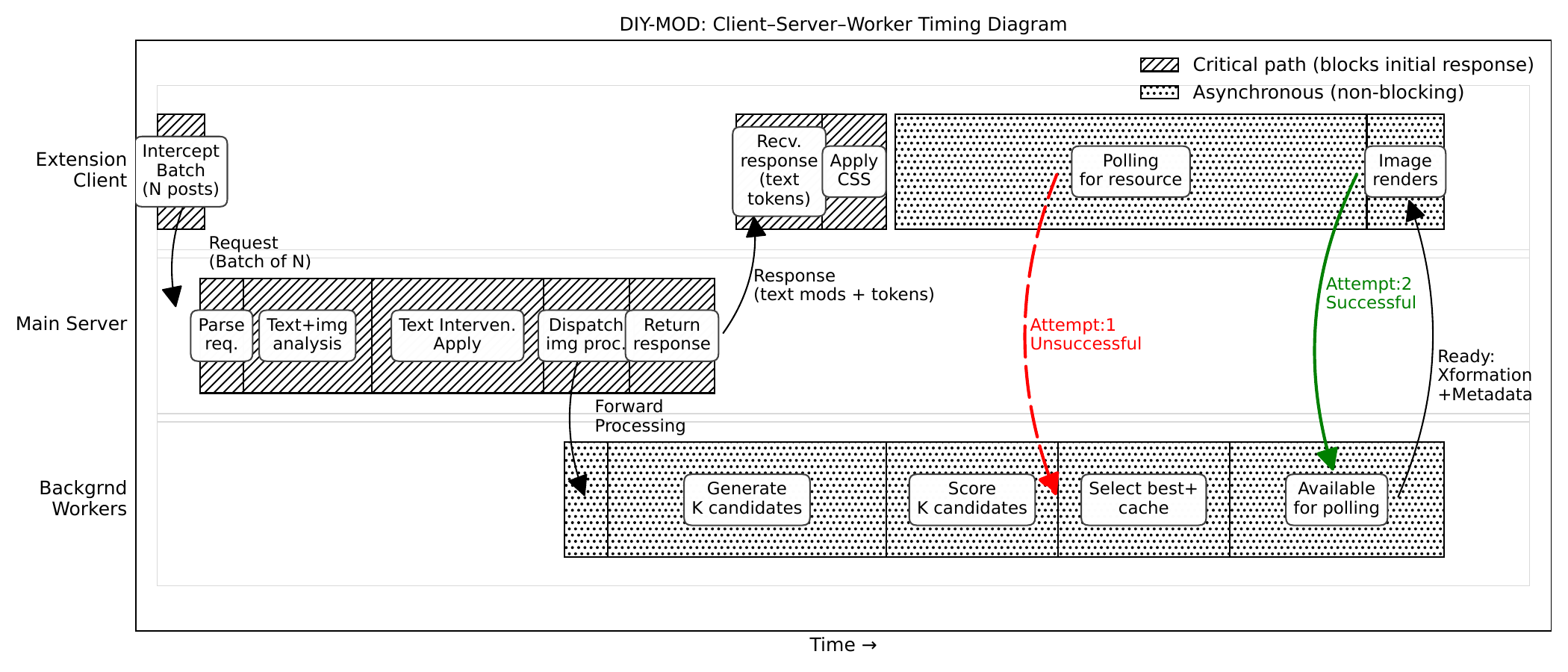}
  }
   \caption{DIY-MOD processing pipeline showing synchronous and asynchronous execution paths.  The critical path (diagonal hatching) completes in 5-15 seconds \new{for a batch of posts}, returning text modifications and polling tokens. Image transformations process asynchronously in background workers (dotted pattern). The client polls for completed transformations—missing on attempts until the content transformation is ready. Previously processed content can return immediately from cache on first poll. \new{Note that, the critical path delay is for a batch of usually 25 posts in reddit. And, this delay only happens when the user first loads/refreshes the feed in browser. For subsequent batches of posts within the browsing session, predictive prefetching (\Cref{appendix:prefetching}) minimizes the delay under reasonable scrolling behavior.}} 
  \label{fig:timing_swimlane}
\end{figure*}

Our primary optimization decouples text and image processing timelines. The system returns text interventions immediately while deferring computationally intensive image transformations to background workers.

When the server receives a batch of posts, it initiates parallel processing through asyncio tasks. The critical path includes text filter matching, text intervention application, and image filter analysis, which complete within 5-10 seconds total. Text modifications return immediately with CSS-based intervention instructions, while images that require transformation receive placeholder tokens. The server then initiates Celery worker processes for the computationally intensive image transformation pipeline. The client then polls for completed image transformations using these tokens.

\noindent Our polling strategy reflects \textbf{two} key insights.\\ \textit{First}, analyzing the DOM structure, we found that Reddit employs lazy loading for images, fetching actual image content only when necessary based on viewport position and network conditions. Our deferred image processing aligns naturally with this behavior. \\ \textit{Second}, the probability that intervention selection has completed for a content increases over time as more candidates are evaluated. Therefore, the client implements an inverse exponential backoff polling strategy, starting with longer intervals that gradually decrease. To prevent server overload from synchronized polling, each content item adds randomized jitter to its polling delay. Posts appearing earlier in the feed receive higher polling priority since users are more likely to encounter them first.

This architecture ensures that users can immediately interact with text content while image transformations process in the background. It is worth noting that, \textit{we do not touch platform advertisements throughout this process}, preserving the platform's native monetization mechanisms.

\subsection{Predictive Prefetching}
\label{appendix:prefetching}
\textit{Predictive prefetching}  leverages the natural pace of browsing. We observed that users typically spend 30-90 seconds engaging with the posts in a single batch, which creates a crucial viewing window to process the next batch before a user scrolls to it.
To exploit this window, our server manipulates the platform's pagination cursor in its response to the client. As a result, when a user has viewed approximately 40\% of the current batch, the platform's front-end is prompted to request the next batch of posts, essentially far earlier than it normally would. 

\diymod{} as usual intercepts this request and sends it to our server, ensuring the processing completes long before the user ever reaches the new content. As a result, under reasonable scrolling behavior, users experience no perceptible delay after the initial page load.

\subsection{Content-Based Caching}
\label{appendix:caching}

\diymod{} implements a \textsc{Redis} based cache to eliminate redundant processing when multiple users encounter the same content or when a page reloads. 
The cache key combines three components:
\begin{equation*}
\begin{aligned}
\mathtt{cache\_key} &=
  \mathtt{hash}(\mathtt{content})
\\ &\quad \oplus \mathtt{hash}(\mathtt{filter\_params})
\\ &\quad \oplus \mathtt{hash}(\mathtt{sensitivity\_level})
\end{aligned}
\end{equation*}

This formulation ensures that identical content processed with identical filter parameters yields the same cached result. The cache operates across all users without compromising privacy since only the anonymous analysis result is stored with no link to user identity.

When viral/popular content spreads across Reddit, the first user to encounter it triggers the full processing pipeline. Subsequent users with similar filters receive instantly cached interventions. This particularly benefits popular posts that appear on \texttt{r/all}, \texttt{r/popular} or trending subreddits where many users may have similar content sensitivities.





\begin{figure*}[]
    \centering
    \begin{subfigure}[b]{0.40\textwidth}
        \includegraphics[width=\textwidth]{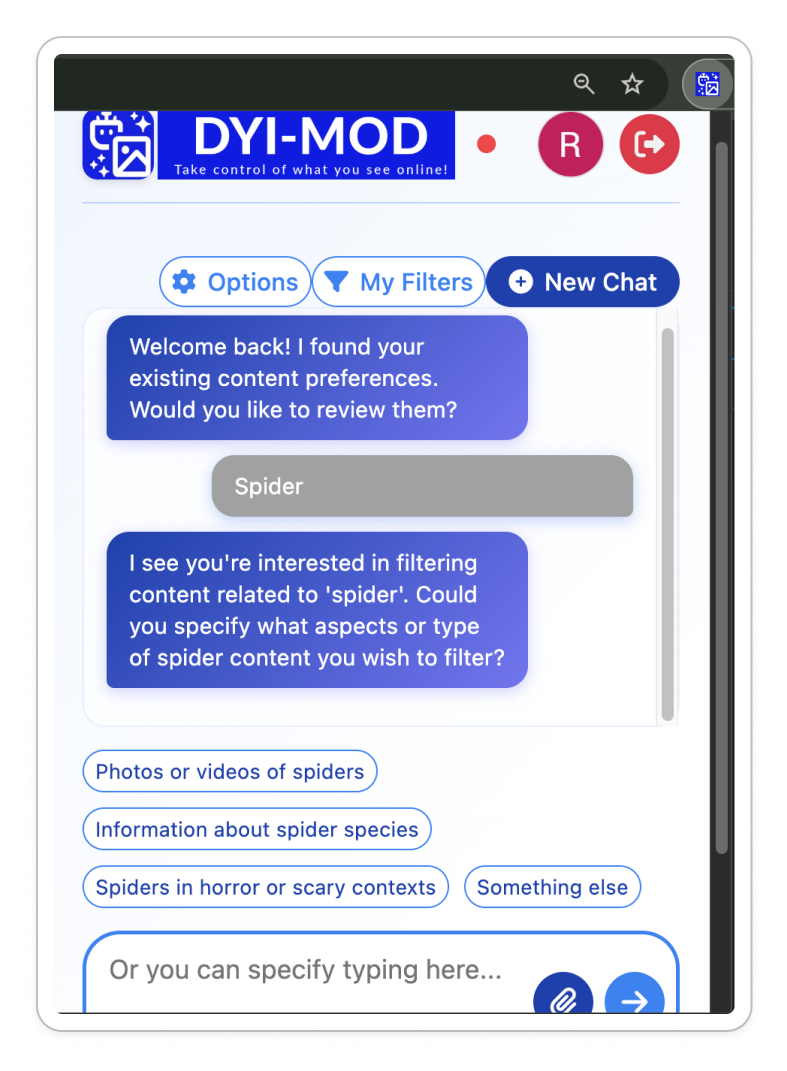}
        \caption{Chat interface}
        \label{fig:chat-interface}
    \end{subfigure}
    \hfill
    \begin{subfigure}[b]{0.40\textwidth}
        \includegraphics[width=\textwidth]{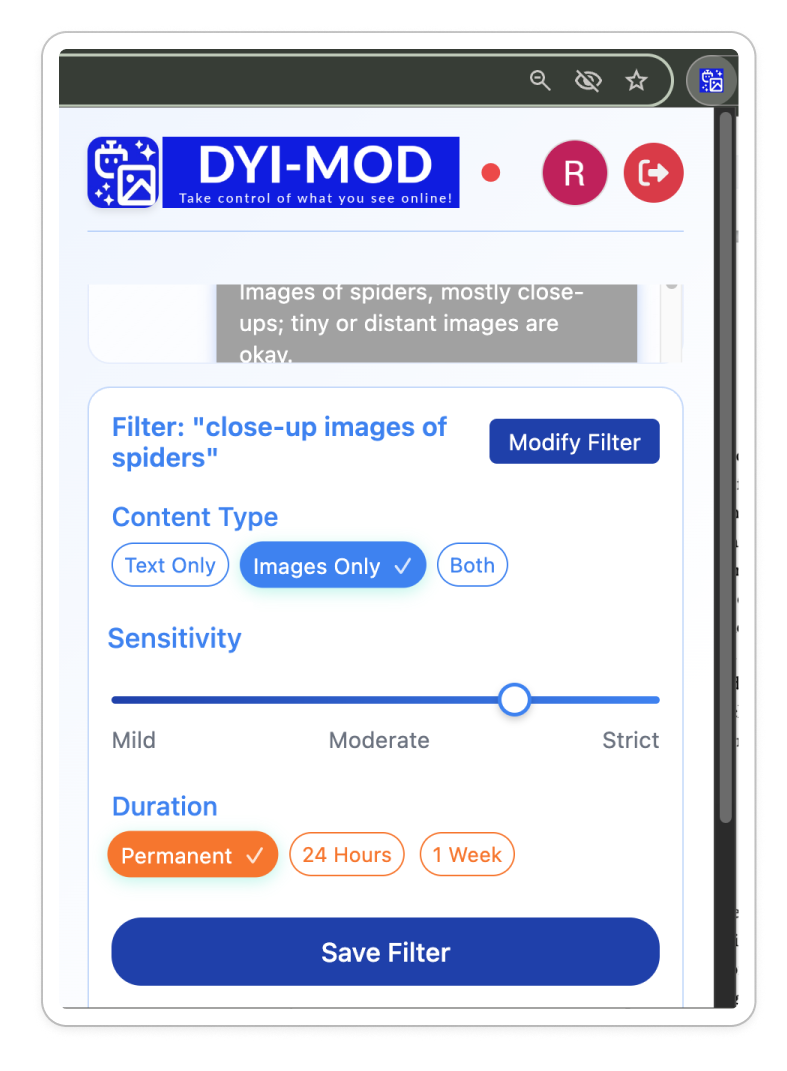}
        \caption{Config interface}
        \label{fig:filter-config}
    \end{subfigure}
    \caption{\diymod{}'s filter creation and configuration interface\newline\textbf{(\ref{fig:chat-interface})} \textbf{Chat interface for filter creation.} \diymod{} engages in conversational grounding to help users specify their content sensitivities. When a user enters a broad term, the system can prompt for clarification about more specific aspects. Users control specification depth through suggested options or free text input. This chat interface opens as a popup when user clicks on the extension icon. \vspace{4pt}
    \newline\textbf{(\ref{fig:filter-config})} \textbf{Filter configuration interface.} After establishing the filter description, users configure three parameters: (1) \textit{Content Type} specifies whether the filter applies to text, images, or both; (2) \textit{Sensitivity Level} indicates the user's distress intensity; and (3) \textit{Duration} sets filter expiration. For brevity, in popup interface we only showed three time presets. Users can modify the filter description anytime using the "Modify Filter" button or add metadata through the Options page for further customization.}
    \label{fig:diy-mod-overview}
\end{figure*}

\section{Filter Creation Interface }\label{appendix:More_About_DIYMOD}\label{appendix:diymod}
Details in Figure \ref{fig:chat-interface} and Figure \ref{fig:filter-config}.

\new{\section{Edit Failures, False Obfuscations and Mitigation Strategies}\label{appendix:failures}}

\new{\subsection{Generation failures and safeguards.} During deployment, we encountered two types of failures in our generation pipeline. First, specific filters occasionally triggered broader VLM categorization: for example, a filter for ``Middle East Conflict'' by one participant matched general conflict imagery or rubble. Second, Gemini's safety guardrails sometimes refused to generate lawful sensitive content such as medical imagery. We investigated this further but observed non-reproducible failures where identical prompts and inputs would succeed on some attempts but fail on others.}

\new{We designed our mitigation strategy to treat both as fail-safe defaults. When generation fails, we first retry, which resolve in most cases. Generating $K=3$ parallel candidates also increases the likelihood that at least one succeeds and scored higher. While we could switch to alternative VLMs upon multiple failure, we prioritized real-time feed rendering over exhaustive fallback strategies.
Even when all candidates fail to generate, we leave the content unchanged but still apply the transparency indicator (\Cref{transparency_modification_indicator}, \Cref{fig:teaser_user_pov}(d)) on top of the original post. This indicator informs users that the content matched one of their filters and was meant to be transformed. Participants reported using this indicator as a personalized warning label—a cue to look away from posts that might be harmful to them. In other words, we err on the side of flagging a post as potentially triggering rather than exposing users to it.}

\new{\subsection{Audit of edit failures and false positives.} To quantify edit failures and false obfuscations, we manually audited posts that our extension marked as modified via the transparency indicator. For each participant, we randomly sampled 5 text posts and 5 image posts, yielding 75 text items and 75 image items. For text, we confirmed via screen recordings that all 75 posts were successfully transformed. For images, we visually matched each transformed image to the original Reddit post and the filter description of the viewer. We found that 66/75 ($\approx88\%$) images were visually transformed, while 9 ($\approx12\%$) were either left unchanged or edited in the wrong region. Among the 66 transformed images, 62 ($\approx93\%$) were correctly transformed according to the viewer's filter description, implying a false-positive transformation rate of roughly 7\% of transformed images.}

\section{Formative Study: Data Analysis and Protection}

We transcribed the interviews verbatim and then deleted the original recordings to avoid retaining identifiable data. Using reflexive thematic analysis~\cite{braun2019reflecting}, we inductively identified the core themes presented in Section \ref{findings:begin} through \ref{findings:end}.  
We pseudonymized transcripts with coded identifiers (e.g., P\_01, P\_12). We stored names and email addresses for recruitment and compensation in a separate encrypted file, deleted them after payment, and never linked them to transcripts. All study data remained on encrypted, university-approved devices and cloud services accessible only to the core research team.

\end{appendix}

\end{document}